\documentclass[sigconf]{acmart}

\AtBeginDocument{%
  }

\copyrightyear{2024}
\acmYear{2024}
\setcopyright{rightsretained}
\acmConference[CHI '24]{Proceedings of the CHI Conference on Human Factors in Computing Systems}{May 11--16, 2024}{Honolulu, HI, USA}
\acmBooktitle{Proceedings of the CHI Conference on Human Factors in Computing Systems (CHI '24), May 11--16, 2024, Honolulu, HI, USA}
\acmDOI{10.1145/3613904.3642669}
\acmISBN{979-8-4007-0330-0/24/05}

\usepackage{caption}
\usepackage[inline]{enumitem}
\usepackage{xcolor} 
\usepackage{multirow}

\newcommand{\systemname}{ARTiST}

\newcommand{\sampletext}[1]{\textit{#1}}
\newcommand{\simplificationWay}[1]{A#1}

\newcommand\myparagraphbold[1]{ \noindent \textbf{#1}}

\newcommand\myheading[1]{ \textbf{#1}}

\newcommand\revblue[1]{{#1}}

\newcommand{\participantquote}[1]{``\textit{#1}''}
\newcommand{\instructionquote}[1]{\textit{#1}}

\begin{document}


\begin{teaserfigure}
  \includegraphics[width=\textwidth]{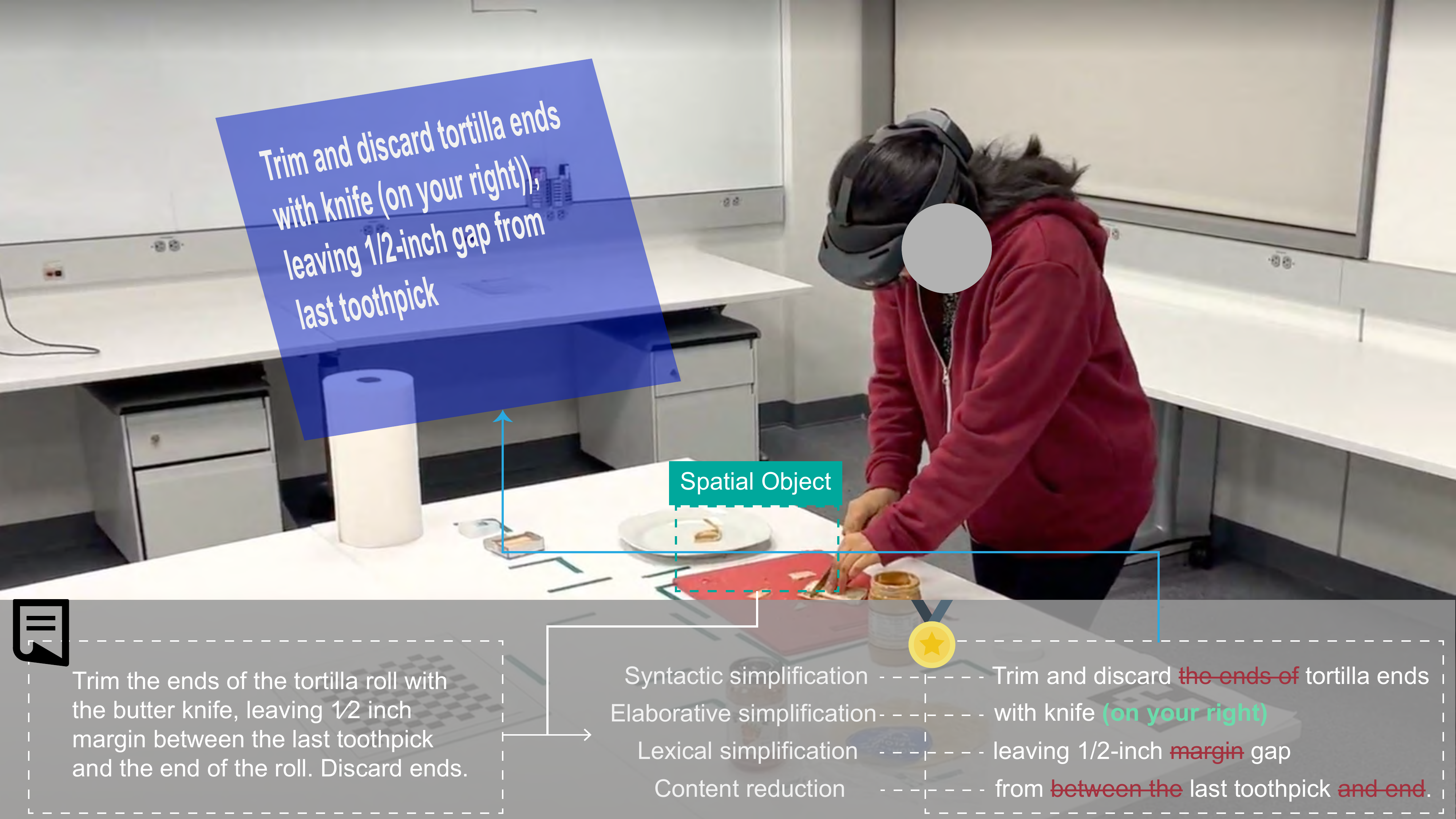}
  \caption{We propose \systemname, a text simplification system that is designed for augmented reality (AR) head-mounted display (HMD) environments. Our system combines the findings from a formative study with a novel few-shot prompting to integrate four established text simplification techniques for AR-specific contexts. The example text shown in the bottom-right corner of the figure has been simplified using our approach. The red text indicates removals whereas the green highlights the addition of spatial information. The resulting simplified text is displayed directly in the HMD.}
  \label{fig:demo-teaser}
\end{teaserfigure}

\title[ARTiST.]{\textit{\systemname}: Automated \textit{T}ext \textit{Si}mplification for \textit{T}ask Guidance in \textit{A}ugmented \textit{R}eality}

\author{Guande Wu}
\email{guandewu@nyu.edu}
\orcid{0000-0002-9244-173X}

\affiliation{%
  \institution{New York University}
  \streetaddress{370 Jay St.}
  \city{Brooklyn}
  \state{New York}
  \country{USA}
  \postcode{11201}
}

\author{Jing Qian}
\orcid{0000-0002-5517-3035}
\email{jq2267@nyu.edu}
\affiliation{%
  \institution{New York University}
  \streetaddress{370 Jay St.}
  \city{Brooklyn}
  \state{New York}
  \country{USA}
  \postcode{11201}
}

\author{Sonia Castelo}
\email{s.castelo@nyu.edu}
\orcid{0000-0001-6881-3006}
\affiliation{%
  \institution{New York University}
  \streetaddress{370 Jay St.}
  \city{Brooklyn}
  \state{New York}
  \country{USA}
  \postcode{11201}
}

\author{Shaoyu Chen}
\orcid{0000-0002-1856-6294}
\email{sc6439@nyu.edu}
\affiliation{%
  \institution{New York University}
  \streetaddress{370 Jay St.}
  \city{Brooklyn}
  \state{New York}
  \country{USA}
  \postcode{11201}
}

\author{Joao Rulff}
\email{jlrulff@nyu.edu}
\orcid{0000-0003-3341-7059}
\affiliation{%
  \institution{New York University}
  \streetaddress{370 Jay St.}
  \city{Brooklyn}
  \state{New York}
  \country{USA}
  \postcode{11201}
}

\author{Claudio Silva}
\email{csilva@nyu.edu}
\orcid{0000-0003-2452-2295}
\affiliation{%
  \institution{New York University}
  \streetaddress{370 Jay St.}
  \city{Brooklyn}
  \state{New York}
  \country{USA}
  \postcode{11201}
}
\renewcommand{\shortauthors}{Wu et al.}


\begin{abstract}
Text presented in augmented reality provides in-situ, real-time information for users. However, this content can be challenging to apprehend quickly when engaging in cognitively demanding AR tasks, especially when it is presented on a head-mounted display. 
We propose \systemname, an automatic text simplification system that uses a few-shot prompt and GPT-3 models to specifically optimize the text length and semantic content for augmented reality. Developed out of a formative study that included seven users and three experts, our system combines a customized error calibration model with a few-shot prompt to integrate the syntactic, lexical, elaborative, and content simplification techniques, and generate simplified AR text for head-worn displays. Results from a 16-user empirical study showed that \systemname~lightens the cognitive load and improves performance significantly over both unmodified text and text modified via traditional methods. 
Our work constitutes a step towards automating the optimization of batch text data for readability and performance in augmented reality.



\end{abstract}



\maketitle

\section{Introduction}
Augmented reality (AR) has evolved into a transformative technology with far-reaching applications across multiple domains, including education \cite{wu2013current, akccayir2017advantages, ibanez2018augmented}, entertainment~\cite{lyu2005arcade, piekarski2002arquake, parekh2020systematic, DBLP:conf/ACMdis/QianZYMCLG021}, collaborative work~\cite{billinghurst1998evaluation, billinghurst2012augmented, ens2019revisiting}, and professional training~\cite{azuma1997survey, billinghurst2012augmented, tatwany2017review, zheng2021applications}. Notably, AR superimposes digital content onto the physical world in real-time to facilitate more efficient task execution~\cite{DBLP:conf/tvx/QianSAB20, xiong2021augmented}. As a result, AR applications have been increasingly employed for task guidance in manufacturing~\cite{zubizarreta2019framework, palmarini2018systematic}, education~\cite{augestad2017educational}, and surgery~\cite{parekh2020systematic}. AR devices have been widely adopted in the manufacturing industry, for example, to reduce reliance on guidance materials or other devices outside of the immediate work environment~\cite{lapointe2020literature}.

A head-mounted display (HMD) is a type of AR device that allows for multi-modal interactions while the user maintains focus on the immediate work environment~\cite{lapointe2020literature, fidalgo2023survey}.
Given their hands-free nature, HMDs are frequently used for text-based task guidance.
However, compared to desktop displays, HMDs have a relatively small field of view (FoV) limiting the amount of text they can display; longer instructions may occlude a user’s view resulting in lower-productivity and higher cognitive load~\cite{DBLP:journals/jcal/BuchnerBK22}, as well as safety risks~\cite{parekh2020systematic}. As a result, AR text-based instructions require optimization for better utility. 

\revblue{Text simplification offers one potential solution. This process has historically been used to reduce the complexity or length of text for users~\cite{siddharthan2014survey} in non-AR settings, making it more readily understandable. To the best of our knowledge, however, there are presently no established methods for adapting and simplifying text for better utility in the specific context of AR.}  Furthermore, applying existing methods to AR raises several concerns. 
Firstly, traditional text simplification methods typically work to facilitate comprehension for individuals with limited reading skills~\cite{shardlow2014survey, carroll1999simplifying}, an audience that may not overlap with the AR userbase.
Secondly, these methods have not been designed or fine-tuned to accommodate AR-specific constraints, such as the small FoV, a restricted display area, or the necessity of users performing physical tasks concurrent with reading~\cite{parekh2020systematic}. Finally, text simplification presents an opportunity to use spatial information AR content by describing a physical object's color, location, or direction.
Textually indicating the location of a physical object can, for example, assist users in AR task execution~\cite{parekh2020systematic, zubizarreta2019framework}.

\revblue{Accordingly, we aim to implement a text simplification system for AR by tailoring the existing methods to the AR context, with the goal of reducing cognitive load on users and improving their task performance.}
To do so, we build on insights from prior work~\cite{chandrasekar1996motivations, shardlow2014survey, siddharthan2014survey, siddharthan2006syntactic} and our own formative study to understand the specific challenges of AR text interfaces as well as their limitations and potentials. The formative study contains three parts: a literature survey, an open-ended exploration, and an expert interview. We found that both participants and experts addressed issues related to long-text-induced reading challenges (e.g., cognitive load) and comprehension. Interviews with participants and experts further elicited three design guidelines that helped build \systemname, an automated text simplification system with few-shot prompting. This system leverages the multi-task capabilities of the large language model (LLM) GPT-3 in combination with our newly formed simplification techniques, eliminating the need for extensively annotated data~\cite{radford2019language}. We crafted prompts based on chain-of-thought principles, considering text simplification in AR~\cite{wei2022chain}. Specifically, \systemname~introduces two novel simplification methods: the ``plan-of-technique'' prompt and ``error-aware'' model calibration, enhancing the effectiveness and reliability of text simplification.

We tested \systemname~via two studies that entailied an empirical evaluation with 16 participants. The first study included task guidance to make pour-over coffee and set up a meeting room according to specific criteria. The second study asked participants to perform video editing on an iPad using AR instructions given through HMD. These studies explore how our proposed system can better benefit participants over the unmodified text and existing methods by assessing related performance metrics, cognitive load, and subjective ratings. The results indicate that \systemname~ significantly improved task guidance performance by reducing the number of errors participants made, increasing the number of steps they correctly memorized, and reducing their cognitive load. 

In summary, we contribute:
\begin{enumerate}
    \item The \systemname, a novel system for text simplification in AR using few-shot prompts and customized GPT-3. This system incorporates chain-of-thought, plan-of-technique, and error-aware calibration to tailor text simplification for AR.
    \item Results and design guidelines from a formative study that includes a literature review, an open-ended exploration with seven participants, and an expert interview with three field experts for text simplification in AR.
    \item A 16-participant empirical evaluation of \systemname~ against baseline and existing methods, which shows that \systemname~ significantly reduces errors and overall cognitive load with similarly higher subjective ratings on text readability, memorability, guidance, and trust among users.
\end{enumerate}

To further support the development of the field, we open-source our implementation\footnote{Code is available at https://github.com/VIDA-NYU/artist}.

\section{Related Work}
Our research draws inspiration from recent studies in task guidance, AR text display, and text simplification. As text simplification establishes the base for our formative study, we discuss it in detail in the following section akin to~\cite{DBLP:conf/chi/ZhengWWM22, DBLP:conf/chi/KayKHM16}.

\subsection{Task Guidance in AR}
\revblue{HMDs have been increasingly used to guide users in procedural tasks, such as cooking~\cite{DBLP:conf/hci/ZhaiCHL20, 10305427}, surgery~\cite{bichlmeier2007laparoscopic}, and maintenance~\cite{zubizarreta2019framework, palmarini2018systematic}.
They allow the seamless and interactive display of task elements and steps in the physical working space~\cite{azuma1997survey, lin2021comparing, thomas2018situated}.
The concept of a task guidance system was initially proposed by Ockerman et al., who envisioned it as a reference and guide for procedural tasks~\cite{ockerman2000review} such as inspection and assembly. For example, AR-based maintenance task guidance and manuals have been developed to display technical instructions interactively in the workspace. Experiments in engine and factory maintenance demonstrate that such manuals can, in comparison to traditional paper manuals,  reduce cognitive load on users and enhance task performance~\cite{DBLP:journals/cii/FiorentinoUGDM14, uva2018evaluating}. Further studies confirm that AR task guidance can lead to a reduced error rate and increased user satisfaction~\cite {tang2003comparative, zheng2015eye}}.
\revblue{Such benefits were also observed in the military~\cite{henderson2009evaluating} and medical applications~\cite{bichlmeier2007laparoscopic}. Furthermore, the visual aspect of textual information can be integrated to increase user engagement and highlight important information~\cite{kim2018does, nijholt2022towards, schmeil2007mara}.} \revblue{However, despite the widespread application of AR task guidance, most existing approaches directly transfer paper-based manuals to AR devices without further customization~\cite{DBLP:journals/cii/FiorentinoUGDM14, kim2018does} or simplification, neglecting the reading challenges posed by AR's FoV. Our approach seeks to bridge this gap by first comprehending the unique challenges of AR as identified in our formative study and then implementing an automatic text simplification method based on the guidelines derived from it.}
\subsection{Text Presentation in AR}
\revblue{Text presentation is an essential function of AR devices; it features in applications that require simple text display through to more robust task guidance. AR text display is especially useful when the text aligns with and supplements the physical environment and the task at hand~\cite{azuma1997survey, DBLP:conf/chi/QianSWHSHT022}.}
\revblue{However, AR text display poses challenges due to the potential for occlusion of the physical environment. Hardware limitations, such as limited FoV, require that AR developers strategically position text within a constrained display are.}
\revblue{To address potential occlusion, solutions that make use of depth information~\cite{wloka1995resolving, holynski2018fast}, inter-frame motion~\cite{lepetit2000semi} and the 2D collision detection algorithms~\cite{breen1996interactive, tian2019occlusion} have been proposed. Though occlusion and collision are not entirely avoidable due to small display areas and the complexity of physical environments~\cite{carmigniani2011augmented}}, we aim to mitigate them by simplifying sentence structure and length. \revblue{Moreover, text presentation can be positively influenced by font selection, placement, and style~\cite{orlosky2014managing, matsuura2019readability}.}
Orlosky \cite{orlosky2014managing} proposed an automated text placement algorithm adaptive to physical and virtual backgrounds. Rzayev et al.further conducted an empirical study on text display type and position in sitting and walking scenarios \cite{rzayev2018reading}, concluding that top-right placement may not be optimal due to increased subjective load and reduced comprehensibility. Matsuura et al. later investigated the readability and legibility of six different Japanese fonts used on HMDs for users who were walking and found that fonts with very thin horizontal lines, or thin horizontal and vertical lines, should not be used on HMDs due to decreased text readability caused by vertical shocks~\cite{matsuura2019readability}. Text coordinates also contribute to the legibility of text presentation in AR applications. Three different coordinates are widely used in AR: body-locked, world-locked, and head-locked \cite{billinghurst1998evaluation}. Body-locked displays are based on the user's body position and adapt to the user’s movements while  walking, making them most suitable for scenarios where users need to walk to perform tasks.

\revblue{Existing methods for text presentation in AR thus focus on font style and layout, but largely overlook the optimization of the text content itself. Our research addresses this present oversight by utilizing text simplification to save display space but preserve essential content.}

\section{Formative Study}
\label{sec:formative}
This study involves three parts: a literature review, an open-ended exploration, and expert interviews to understand the needs around text simplification in AR.
We wish to explore the following aspects of text simplification:

\begin{enumerate}[start=1,label={[\bfseries RQ\arabic*]}]
    
    \item[\textbf{[RQ1]}] \label{req:RQ_1}  Which text simplification methods from the field of natural language processing (NLP) can be effectively applied in an AR context?
    \item[\textbf{[RQ2]}] \label{req:RQ_2} Would text simplification improve comprehension, per its benefit for low-literacy readers in traditional platforms?
    \item[\textbf{[RQ3]}] \label{req:RQ_3} Can text simplification lead to increased user satisfaction, and hence to a more positive AR experience overall? 
\end{enumerate}



\subsection{Part I: Survey of Text Simplification}
To address~\ref{req:RQ_1}, we initiated a comprehensive review of existing literature on traditional text simplification; our goal was to identify NLP techniques that might be useful for AR applications. Our review commenced with an in-depth examination of three seminal survey papers \cite{siddharthan2006syntactic, shardlow2014survey, al2021automated}. We extended our scope by traversing both the references cited in these papers and consequent citations of them to gain an encompassing understanding of current methodological approaches. Subsequently, we identified four NLP techniques pertinent to our inquiry: content reduction (\textbf{A1}), syntactic simplification (\textbf{A2}), lexical simplification (\textbf{A3}), and elaborative simplification (\textbf{A4}). We describe the four techniques below. 

\revblue{
\subsubsection{A1: Content reduction} Content reduction in text simplification aims to achieve clarity and conciseness by eliminating or restructuring non-essential elements without altering the core message \cite{nisioi2017exploring}. Strategies include removing non-essential information, shortening sentences, and eliminating repetition. This technique is particularly beneficial in constrained display environments like those of AR, where succinct, clear text enhances user interaction and comprehension \cite{bower2014augmented}.
}

\subsubsection{A2: Syntactic simplification}
Syntactic simplification involves rephrasing complex grammatical structures into simpler ones while still retaining the original meaning~\cite{siddharthan2006syntactic, DBLP:conf/ijcnlp/ScartonATMS17}. Existing methodologies often target specific complex linguistic features such as coordination, subordination, relative clauses, passive constructions, and extended sentence lengths \cite{siddharthan2006syntactic, al2021automated}. 

\subsubsection{A3: Lexical simplification}
Lexical complexity often arises from the use of intricate words and phrases. To mitigate this, one widely employed strategy is to replace complex lexical items with simpler synonyms. This form of lexical simplification has seen extensive application in the context of second-language learning, primarily because it aids in comprehension and vocabulary acquisition for learners who may not be familiar with advanced or specialized terminology \cite{paetzold2017survey}.

\subsubsection{A4: Elaborative simplification} 
\revblue{Elaborative simplification entails providing explanations of complex concepts.} This technique is prevalent in professional textbooks, which frequently encompass specialized or technical subject matters~\cite{keil2020augmented}. In AR, spatial information becomes increasingly critical for user comprehension and task performance. Consequently, elaborative simplification can be especially beneficial for clarifying spatial metrics and locations. Spatial metrics refer to numerical measurements, such as distances or sizes denoted in units like inches or centimeters. These metrics often need to be elaborated to provide context or improve comprehension. Similarly, spatial locations, which may involve GPS coordinates or relational positioning (e.g., "next to," "above," "beneath"), can be clarified through elaborative simplification to facilitate user orientation and task execution in AR environments~\cite{hornacek2022spatial}.

\subsubsection{Target application and users}
Traditional text simplification techniques are normally targeted at non-native speakers or people with cognitive or literacy limitations, e.g., autism~\cite{dattolo2017accessible, yaneva2015easy, evans2014evaluation}, aphasia~\cite{obiorah2021designing, canning2000cohesive, carroll1998practical}, dyslexia~\cite{rello2013frequent, rello2013simplify, rello2017present, hmida2018assisted, gala2016reducing, ivanova2017ontology}, hearing impairment~\cite{alonzo2020automatic, alonzo2022dataset, alonzo2022methods, alonzo2022use, vettori2011supporting} and language learners \cite{lotherington1993simplified}. 
\revblue{The associated benefits are largely attributed to simplified grammar structures and the use of common words, which can significantly reduce information processing time.}

\revblue{In a similar vein, AR users may encounter reduced reading capability due to the challenges associated with the AR setting.} Studies have demonstrated that AR users experience reduced reading speed~\cite{rau2018speed}, lower comprehension \cite{gabbard2006effects, gabbard2007active, bower2014augmented}, and increased cognitive load~\cite{dunleavy2014augmented}. For instance, Rau et al. report that readers' response time in AR is longer than that associated with desktop reading~\cite{rau2018speed}. Hardware limitations comprise a major contributing factor, impacting refresh rate, resolution, and FoV, and ultimately impeding text display due to registration errors~\cite{macintyre2002estimating, holloway1997registration}, extra latency, and visual artifacts. Moreover, users' rapid movements and the surrounding open environment can result in unstable text displays. Prolonged use of AR devices may also lead to eye strain and fatigue due to constant accommodation and vergence adjustments, making reading more challenging than on traditional displays. Finally, AR displays often overlay digital information on real-world information, affecting reading comprehension and focus.

Yet, the main reason for the reduced legibility of AR text is the user's elevated cognitive load in the immersive environment; evidence shows that AR users can experience high pressure and increased cognitive load \cite{dunleavy2014augmented}.

\revblue{Accordingly, drawing inspiration from the research elaborated above, we \textit{aim to investigate whether text simplification techniques can benefit users in AR environments and mitigate the challenges described}.}

\subsection{Part II: Open-Ended Exploration} 
\revblue{To explore the effectiveness of the four previously identified text simplification techniques (\textbf{A1-4}) in AR and further investigate \ref{req:RQ_2}, we conducted an open-ended exploration with seven participants. According to the literature, the simplification techniques (\textbf{A1-4}) can improve comprehension in paper-based reading. However, these techniques may need to be modified for AR-specific challenges and their benefits in AR require further investigation. Therefore, the open-ended exploration aimed to assess how text simplification techniques might be revised for this context.}

\subsubsection{Participants}
Seven participants (four male and three female) were recruited from a school mailing list to experience text simplification in AR. Three out of the seven are native English speakers. All participants have some prior experience with AR (2/7 are frequent users, 4/7 are occasional users, 1/7 is VR-only). 

\subsubsection{Tasks}
\revblue{The open-ended exploration involves two tasks: cooking and gem-hunting. These tasks were selected for being common and applicable to AR scenarios~\cite{DBLP:conf/chi/KumaranKMBGH23, DBLP:conf/hci/ZhaiCHL20, DBLP:journals/sensors/MajilYY22}.} In the cooking task, participants used the AR system to make a pinwheel sandwich. The AR interface showed step-by-step instructions for ingredient preparation, assembly, and cooking~\cite{bower2014augmented}. These instructions are adapted from a wikiHow article on how to make a pinwheel.~\footnote{https://www.wikihow.com/Make-a-Pinwheel} For the gem-hunting task, participants followed clues displayed on the AR device to find a gem hidden in a room. Clues included puzzles, patterns, and spatial information. The task manual is derived from a party game website.

\subsubsection{Method}
Since each task contains multiple steps, the original and simplified text for each step were displayed side-by-side to participants. . \revblue{Text simplification was manually performed based on the principles of the four existing techniques taken from the literature~(i.e., \textit{content reduction}, \textit{syntactic simplification}, \textit{lexical simplification}, and \textit{elaborative simplification}), with each simplification technique being used an equal number of times.} Since this is an exploratory study, quantitative data is not collected. Participants are asked to think aloud while performing their AR tasks. A semi-structured interview collects participants' thoughts on text simplification in AR, its potential, and its challenges.

\subsubsection{Procedure}
Initially, participants completed both tasks using the original text instructions. They shared any challenges they faced in understanding the AR interface. Next, simplified versions of the text were presented. Participants compared and evaluated readability and comprehension. On average, the exploration lasted about one hour. We coded our interview notes and think-aloud notes and summarized participant feedback on the four text simplification techniques. \revblue{The open-ended study was supervised by the university-approved IRB, and participants were compensated at an hourly rate of \$20.}

\subsubsection{Results}
\revblue{
During the study, we found text content and semantics affect the reading experience in AR.}

\myheading{Text length in AR.}
Text in an AR environment introduces unique challenges that are not present in traditional display mediums. Users can scroll or zoom to manage lengthy texts in conventional formats; these interactions are more challenging to execute in the AR setting~\cite{mulloni2010zooming, buschel2019investigating}.
Occlusion and visual clutter are some of the issues pointed out by our participants (P2), who mentioned, \participantquote{The displayed text takes up too much space and occludes the table.}
Lengthy text segments also distract users’ attention away from physical tasks. P3 found it challenging to focus on the task of \instructionquote{sliding floss under the tortilla, perpendicular to the length of the roll,} due to the distracting nature of the extended text. These distractions sometimes pose safety risks: P7 was at risk of cutting their finger while engrossed in reading.
\revblue{Furthermore, text length negatively impacts how well information is retained as processing time increases with longer text segments~\cite{genzel2002entropy, keller2004entropy}. This was evident in the gem-finding task, where P4 and P5 forgot a crucial step that they had been given earlier after reading a lengthy sentence.} Text length thus requires careful design in AR.

\myheading{Feedback on AR text simplification techniques}
All participants agreed that \textit{content reduction} is beneficial in AR. For instance, they found the sentence, \instructionquote{Roll the tortilla into a log shape,} more effective than the original text: \instructionquote{Roll the tortilla from one end to the other into a log shape.} Most participants mentioned that adding a clause to further explain text may not be necessary (preferring \textit{syntactic simplification}). When asked about replacing complex words with simpler ones (\textit{lexical simplification}), most participants (6/7) did not indicate word complexity as an issue. For example, the word \instructionquote{perpendicular} was not found to be more opaque than \instructionquote{at the right angle too,} and most of the participants (5/7) preferred \instructionquote{perpendicular} because it was shorter (4/7) and more precise (3/7). In addition, most participants (6/7) expressed that added details (\textit{elaborative simplification}) were unnecessary. P6 said that \participantquote{the `which' clause is verbose and takes up too much space}~(referring to the instruction \instructionquote{use the keys to unlock the first drawer below the desk, which should be located to your right}). For spatial elaboration, most (6/7) found it helpful when the reference object was present in the scene. P6 remarked that indicating, \participantquote{`finger size' helps me make a quick estimate of the size.} \revblue{P3 commented that indicating a spatial location in the text is helpful, and a majority of participants (4/7) said that spatial information can complement AR spatial indicators such as bounding boxes or virtual arrows in the scene. }

\subsection{Part III: Expert Interview}
\revblue{To verify the initial insights gained from the literature review and the open-ended exploration, we further conducted a semi-structured interview with three experts from the industry. All interviewed experts possess extensive experience with AR task guidance systems. Our objective in these interviews was to address \ref{req:RQ_3} by eliciting their insights on text simplification for AR and exploring potential usage scenarios.}
\label{sec:expert}

\subsubsection{Expert background}
Each of the three experts (E1-E3) interviewed has over three years of professional experience in AR interface development.
\begin{itemize}
    \item E1 is an AR interface designer at a research and development (R\&D) company that is currently working on a HoloLens application to support field surgery. E1's users are primarily skilled professionals such as teachers and emergency medical technicians (EMTs) who use AR devices to instruct them as they identify and treat injuries such as gunshot wounds.
    \item E2 is an interface developer at a document solution corporation. E2 collaborated with engine mechanics to develop a HoloLens-based instructional application for displaying engine maintenance documents.
    \item  E3 is an AR/VR researcher with top-tier publications and extensive experience in HoloLens application development. E3 has developed AR applications for everyday tasks such as cooking for non-professional users.

\end{itemize}

\subsubsection{Method}
\revblue{The interview addressed the experts’ backgrounds and experiences, the challenges of AR text interface design, their assessment of the need for text simplification, and the potential benefits and drawbacks associated with it. Additionally, we presented the four commonly used text simplification methods and solicited their opinions on them.}

\subsubsection{Results}
We describe the results in the following subsections. 

\myheading{Benefits of text simplification in AR} All experts recognized the need to simplify text in AR. They believed this would reduce user impatience and the likelihood of mistakes. E2 said that mechanics, for instance, might be habituated to how they perform a specific task and so rush through it without noticing updates to the process. When the related instructional text is simplified, however, they are more likely to read the instructions. E2 explained: \participantquote{One of the things that happens is the procedure changes. Users can easily go on a routine and assume they know how to do it without actually reading the instructions.} E1 and E3 also mention that the simplified text could reduce the cognitive load and mitigate user anxiety, another set of benefits. For example, E3 indicated that: \participantquote{Reading the long text may make the users anxious,} but this may not be the case for shorter pieces of text.

All experts indicated that simplified text reduces the chance of visual occlusion. Object occlusion (virtual objects being blocked visually by physical objects~\cite{tian2019occlusion}) is one such instance of this. This leads to users being unable or only partially able to read the AR text, causing frustration and diminished performance. E2 mentioned that: \participantquote{(Sometimes in engine maintenance) we're gonna have a wall full of the tools, (and sometimes) we are gonna have an engine in front of you, and (so) finding someplace in the visual display is gonna be a challenge.} E3 also mentioned that shortening and simplifying text could reduce occlusion. 

Both E2 and E3 mentioned that shorter text facilitates the AR reading experience since zooming or scaling long sections of text while reading on an HMD is challenging. E2 emphasized that: \participantquote{None of the users liked pinching and zooming,} highlighting the need for methods that do not require additional interactions.

Finally, the experts mentioned the tremendous opportunity to use text simplification as a way to help automate the conversion of text from traditional digital media (e.g., PDF) to AR. All experts conveyed that the process of creating text instructions for AR is still sub-optimal and requires extra labor. E2 stated: \participantquote{All the documents we work with start as PDF or Word documents. We basically output them to AR (devices)}. In contrast, E1 and E3 mentioned the need to make modifications to the text displayed in AR. For example, E1 attempts to shrink text or split long sections of text into multiple steps to make them shorter, saying: \participantquote{We try to keep the words as quick, punchy, and actionable as possible.} E3 also mentioned adjusting font sizes and colors to improve legibility in the AR environment. 
\revblue{Although full-text automation involves fitting text to the AR scene with different formats, styles, or colors, E2 pointed out that automated text simplification would still be useful as the current manual approach requires expertise that novice workers may not possess. In addition, it is not feasible to manually revise all text when new sections are added regularly.}



\myheading{Challenges in text simplification in AR} 
Current AR applications lack automated solutions and established practices for text simplification (E1-3). All experts concur that manual text revision is impractical due to the constant influx of new text and the absence of a standard framework for AR text readability. This drives home the need for automated methods to adapt existing documents for presentation in AR. However, text simplification for AR poses the following challenges, and current methods are not directly applicable (E1-3).

\begin{itemize}
\item All experts raised concern over avoiding accidental changes to meaning during text simplification. This concern is unique to AR because users perform physical actions \instructionquote{live} from textual instructions. E2 elaborated: \participantquote{When working with mechanical systems in real-world scenarios, failure to follow instructions accurately could lead to catastrophic consequences,} highlighting the importance of retaining the integrity of the original text's meaning.

\item Removing duplicated content is crucial in AR given that such redundancies could increase cognitive load for AR users who are already tasked with interpreting and acting upon visual overlays. All experts agreed that elaborative simplification should weigh toward eliminating redundancies rather than adding explanatory details, which is traditional in conventional text simplification.

\item Traditional text simplification techniques must be re-adapted for AR (E1-3) as they are primarily geared toward enhancing readability for low-literacy individuals and do not address the attention constraints, high cognitive load, and FoV issues typical in AR. Therefore, the development of an AR-specific text simplification tool presents a challenging yet vital task, as it must harmonize these design goals to suit the unique demands of AR settings.

\end{itemize}


\subsubsection{Feedback on existing text simplification techniques}
\begin{table}[ht]
\centering
\begin{tabular}{|l|l|l|l|l|}
\hline
  \textbf{ID} &      \textbf{Technique}                 & \textbf{E1}                      & \textbf{E2}                      & \textbf{E3}                      \\ \hline
 \simplificationWay{1} & Content reduction        &                          & \checkmark &                          \\ \hline
\simplificationWay{2} & Syntactic simplification & \checkmark & \checkmark & \checkmark  \\ \hline
\simplificationWay{3} & Lexical simplification   & \checkmark  &                          &                          \\ \hline
\simplificationWay{4} & Elaborative simplification     &                        \checkmark  & \checkmark  & \checkmark                    \\ \hline
\end{tabular}
\caption{\revblue{Expert (E1-3) feedback on simplification techniques (A1-4). A check indicates that the expert assesses that the given technique would be useful for AR.}}
\label{tab:methods_feedback}
\end{table}

 The table~\ref{tab:methods_feedback} summarizes the traditional simplification techniques our experts use in their everyday work. All experts do manual~\textit{content reduction} when creating AR instructions. E3 employs lexical simplification with the aim of retaining the text's original meaning. All experts agree that simplifying syntax, length, and grammar is beneficial for AR interfaces. However, the use of elaborative simplification needs more scrutiny in AR settings as \participantquote{the subtle balance between the content and text length must be considered}(E1, E2, E3). For example, E1 mentioned that engine maintenance manuals often include explanations of different engine parts that may be unfamiliar to users, and such explanations should not be removed. E2 brought up that both object and numeric elaboration can be beneficial when users need to quickly identify numerous targets in AR. Elaborating AR text to describe objects in the scene is one potential application. 
 \revblue{E2 explained that using a reference object that is similar in size to the dimensions given in the text (when the object is visible) would facilitate spatial awareness.}
For example, E2 said that a phrase like \instructionquote{Move the handle to seven inches to left} can be elaborated as \instructionquote{Move the gear to seven inches left, or the length of a screwdriver}. Again, as the experts mention, consideration needs to be given to balancing text length against the need for additional content in AR.

\subsection{Design Guidelines and Updated Simplification Techniques}
\label{sec:dg}
\subsubsection{Design guidelines}
\revblue{Through summarizing the literature survey, the open-ended exploration, and insights shared by our AR experts, we derived design guidelines and updated the four selected simplification techniques for AR task guidance.}

\begin{enumerate}[start=1,label={[\bfseries DG\arabic*]}]
\item[\textbf{[DG1]}] \label{design_guideline:DG_1} \textbf{Meaning preservation is paramount in text simplification.} Preserving the original text meaning~\cite{siddharthan2014survey, beigman2004text}) is the main objective when applying text simplification techniques. This finding is in line with both our interview sessions and observations. Since almost all simplification techniques may compromise original meaning~\cite{nisioi2017exploring, siddharthan2006syntactic}, it is essential that any substituted words convey the same meaning as their original counterparts~\cite{devlin1998use}.

\item[\textbf{[DG2]}] \label{design_guideline:DG_2} \textbf{\revblue{Text simplification must consider both AR-specific challenges, such as issues with FoV and cognitive load, while exploring AR-specific opportunities.}}
\revblue{Traditional text simplification techniques (e.g., syntactic simplification, lexical simplification, etc.) do not address challenges associated with AR devices, such as reading the overlayed text while doing a physical task, the constraints of a small FoV, and users' increased cognitive load while completing a task.} Minimizing the display space required to render text reduces the chance of visual occlusion while optimizing syntactic structures reduces cognitive load.

\item[\textbf{[DG3]}] \label{design_guideline:DG_3} \textbf{Text simplification in AR should give priority to text length over grammatical correctness} 
Traditional text simplification techniques usually prioritize grammatical correctness \cite{siddharthan2006syntactic, al2021automated}. \revblue{However, we find that priority should instead be given to text length and clarity in AR. This was gleaned from the open-ended exploration, where participants expressed the need to minimize occlusion caused by text length and indicated that less strict grammar did not notably affect their comprehension if meaning was preserved. Further expert interviews supported the assessment that AR users tend to skim lengthy texts, not paying strict attention to grammatical correctness.}

\end{enumerate}

\subsubsection{Updated simplification techniques (A1-4)}
\revblue{Based on our findings, we update the four simplification techniques to fit users' needs in the AR context. We discuss the benefits and address discrepancies within the experts' feedback below.}

\myparagraphbold{A1: Content reduction}
\revblue{We found that content reduction is beneficial in AR, as both the literature review and experts suggest. However, removed content may contain important task instructions, and its absence may alter the original meaning~\ref{design_guideline:DG_1}. Furthermore, as suggested by \ref{design_guideline:DG_3} and observation of the open-ended exploration, prepositions and pronouns can be cut for more concise.}

\myparagraphbold{A2: Syntactic simplification}
\revblue{The results from the formative study support syntactic simplification as beneficial in AR contexts, given that complex grammatical structures can consume user attention. However, as with content reduction, syntactic simplification may alter the original meaning~\cite{siddharthan2006syntactic}, necessitating adherence to~\ref{design_guideline:DG_1}. Furthermore, simplified grammatical structure can result in overall longer text, contradicting~\ref{design_guideline:DG_3}. To mitigate this, syntactic simplification should be applied only when it does not increase the number of lines of the displayed text, as addressed by E3.}

\myparagraphbold{A3: Lexical simplification}
\revblue{
Lexically simplified phrases may deviate from original meanings and lengthen the text, conflicting with~\ref{design_guideline:DG_1} and~\ref{design_guideline:DG_3}. To address this, we propose two constraints for lexical simplification: Firstly, it should not alter task-related terms, and, secondly, it should not increase the number of lines of text.
}

\myparagraphbold{A4: Elaborative simplification}
Elaborative simplification elicited nuanced opinions from the experts. In the NLP literature, elaborative simplification is described as benefiting second-language learners by elucidating abstract terms. However, as noted by E1-3, explaining terms may not benefit AR users and will likely lead to increased text length (contrary to \ref{design_guideline:DG_3}). Therefore, common-sense explanations and explanations of background knowledge should be excluded from elaborative simplification to support concision. However, E1-3’s feedback indicates that elaboration of the spatial context and numerical measures offers greater utility within the AR context. For instance, the user can benefit from spatial positional information such as \instructionquote{the cup on your left} when multiple cups are present. Additionally, when conveying numeric measures (e.g., \instructionquote{seven inches}), experts advised elaborating by referencing the size of objects already present within the scene, such as the diameter of a plate. By incorporating spatial context and numerical measure, elaborative simplification can be adapted within AR to adhere to \ref{design_guideline:DG_2} and enhance task performance.

\section{\systemname~System}

\begin{figure*}[ht]
    \centering
    \includegraphics[width=\linewidth]{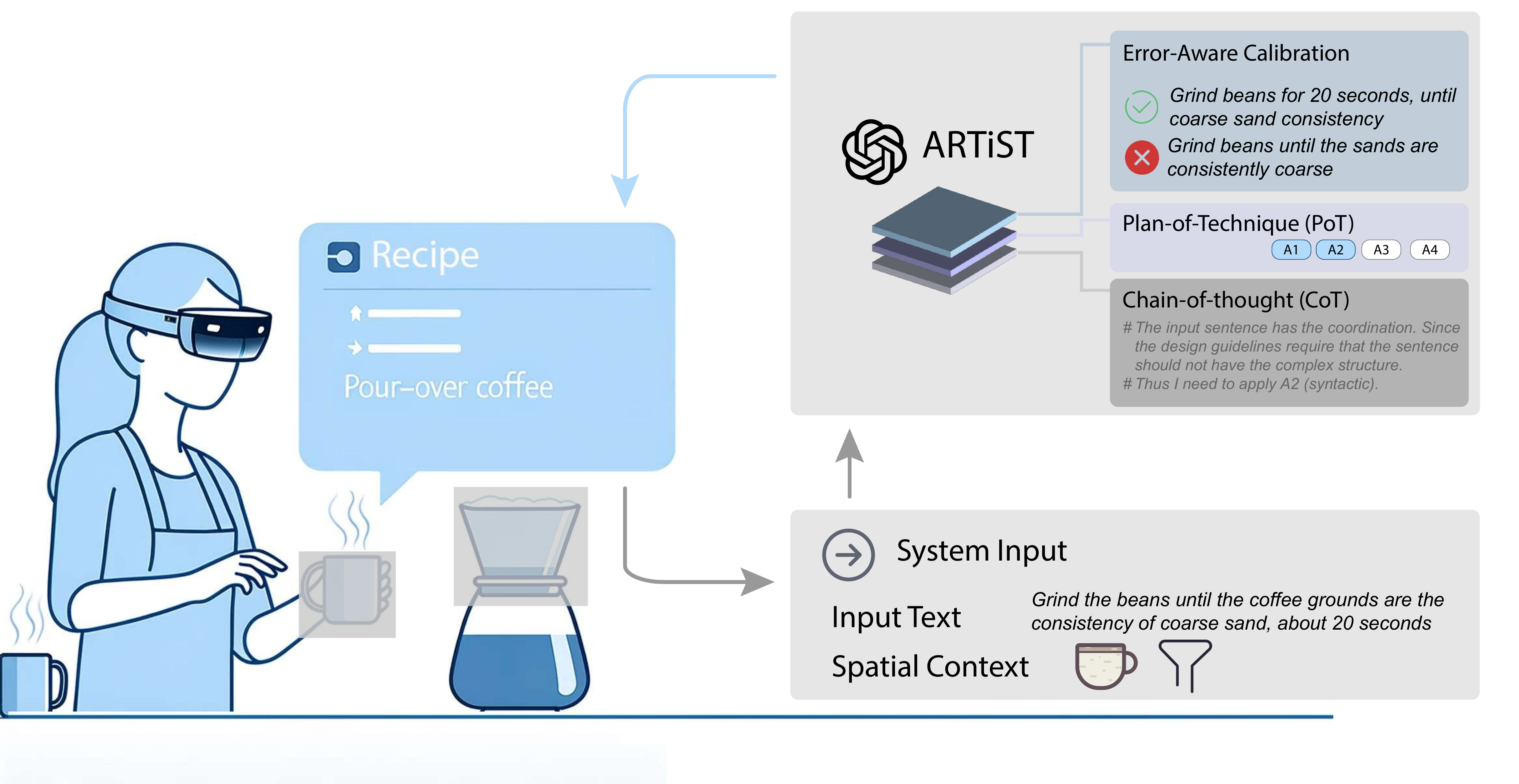}
    \caption{\revblue{Method overview: \systemname~uses OpenAI's GPT-3 model, prompted with chain-of-thought and technique-as-plan methods, to generate simplified text candidates. The candidates are calibrated to reduce the likelihood of potential errors. The resulting simplified text is then displayed within a HoloLens 2 application. The spatial context is captured by detecting the objects in the scene to support the elaborative simplification.}}
      \label{fig:method}

\end{figure*}
\revblue{In this section, we describe the design of \systemname, which has been developed using the updated simplification techniques (Table \ref{tab:methods_feedback}) and design guidelines  derived from the formative study.}

\systemname~employs three novel methods, shown in Figure~\ref{fig:method}, to customize the GPT-3 model to stably output the desired simplification results. 
These include utilizing the chain-of-thought method to enhance GPT-3's reasoning capabilities and the plan-of-technique method for selecting the most appropriate techniques from \textbf{A1-4}~(\ref{design_guideline:DG_2} and \ref{design_guideline:DG_3}). Additionally, \systemname~implements error-aware calibration to ensure the preservation of the original text's meaning(\ref{design_guideline:DG_1}).


\subsection{Plan-of-Technique Prompting}
\begin{figure*}[ht]
  \centering
  \includegraphics[width=\linewidth]{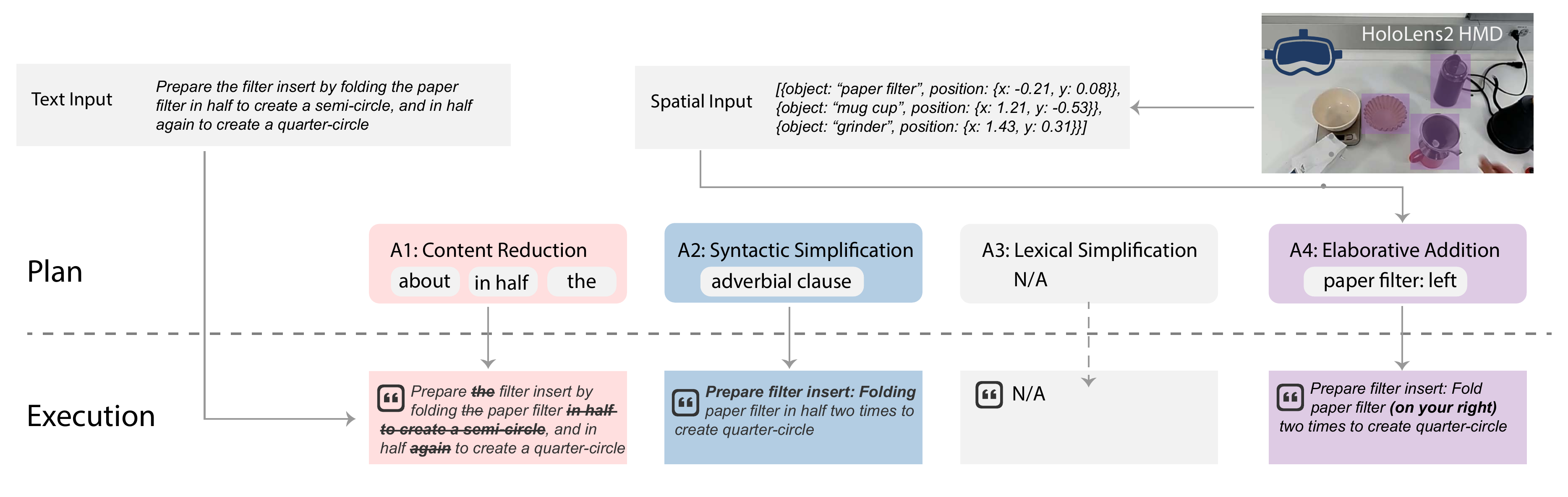}
  \caption{\revblue{Plan-of-technique: The input text and the spatial context are fed into the LLM, which first generates a plan of the simplification techniques. The techniques will be sequentially applied to the input text to generate the final simplified text.}}
    \label{fig:pot}

\end{figure*}
{The plan-of-technique method is designed to structure the simplification process through a plan of different simplification techniques (\textbf{A1-4}). These techniques guide the GPT-3 model in executing the simplification as intended.
\revblue{This planning-and-execution model has been widely adopted in code generation \cite{DBLP:journals/corr/abs-2303-05510}, open-world agents \cite{DBLP:journals/corr/abs-2302-01560}, and robotics \cite{DBLP:journals/corr/abs-2209-11302} for controllable and stable outputs.}
\revblue{Figure~\ref{fig:pot} shows how input texts and the spatial context are fed into GPT-3 to generate the simplification plan. }

\revblue{Step-by-step execution ensures that all necessary simplification techniques can be applied. Our preliminary experiments reveal that GPT-3 sometimes forgets the techniques and design guidelines. One explanation for this is that LLMs like GPT-3 are typically trained on generic corpora without access to specialized design guidelines. Our plan-of-technique thus decomposes the simplification process into different simplification steps, mitigating forgetfulness. Moreover, such a structured pipeline can elicit the GPT-3's multi-hop reasoning capability shown in other NLP tasks~\cite{DBLP:conf/emnlp/Adolphs0USW22, DBLP:conf/emnlp/0001KARSW22, DBLP:conf/eacl/SunRR23}.} 

\revblue{In text simplification, multiple simplification techniques can sometimes conflict with each other and require multi-hop reasoning to resolve. For example, elaborative simplification (\textbf{A4}) may conflict with content reduction (\textbf{A1}). The plan-of-technique guides GPT-3 to consider the different techniques before executing the actual simplification actions, thereby reducing potential conflicts.}
 


\subsection{Chain-of-Thought Prompting }
\revblue{Chain-of-thought prompting is used to further enhance GPT-3's multi-hop reasoning capabilities and resolve potential technique conflicts. In few-shot prompting, a series of exemplars are created to instruct GPT-3 on how to generate the desired output based on the input text. Chain-of-thought augments the exemplars with intermediate reasoning steps, leading to the final output~\cite{wei2022chain}. }
Drawing upon the proven efficacy of chain-of-thought's applications in diverse fields \cite{DBLP:conf/emnlp/Wang0S22, DBLP:conf/eacl/KimJJCY23, si-etal-2023-getting}, we incorporate chain-of-thought into both the planning and execution phases of the plan-of-technique method. This decision aligns closely with~\ref{design_guideline:DG_2} and~\ref{design_guideline:DG_3}, which stress the importance of adaptively applying traditional text simplification techniques (\textbf{A1-4}) to cater to AR-specific needs. 

We use an example to show how the chain-of-thought method supports the plan generation in the plan-of-technique method. To simplify the sentence, \sampletext{Grab a pair of 10 to 12 lb (4.5 to 5.4 kg) dumbbells and lie on your back with your arms behind you and your legs extended and raised to a 45-degree angle}, we prompt GPT-3 to generate the thoughts about the input text's applicability to AR context. GPT-3 identifies the sentence as overly lengthy, containing more than three phrases, and thus includes syntactic simplification in its simplification plan. 
The plan involves three steps of syntactic simplification: \begin{enumerate*}
\item splitting the sentence at the first \sampletext{and} because the length of the two joined clauses is too long;
\item splitting the sentence at the second \sampletext{and} for the same reason.
\item Adjusting the passive voice in \instructionquote{your legs extended and raised} for better readability.
\end{enumerate*} 
After generating the plan, we continue to prompt GPT-3 to apply the simplification techniques outlined in the plan, yielding the result: \instructionquote{Grab a pair of 10 to 12 lb (4.5 to 5.4 kg) dumbbells. Lie on your back with your arms behind you. Extend your legs and raise them to a 45-degree angle.}


\subsection{Error-Aware Model Calibration}
\begin{figure*}[ht]
    \centering
    \includegraphics[width=0.8\linewidth]{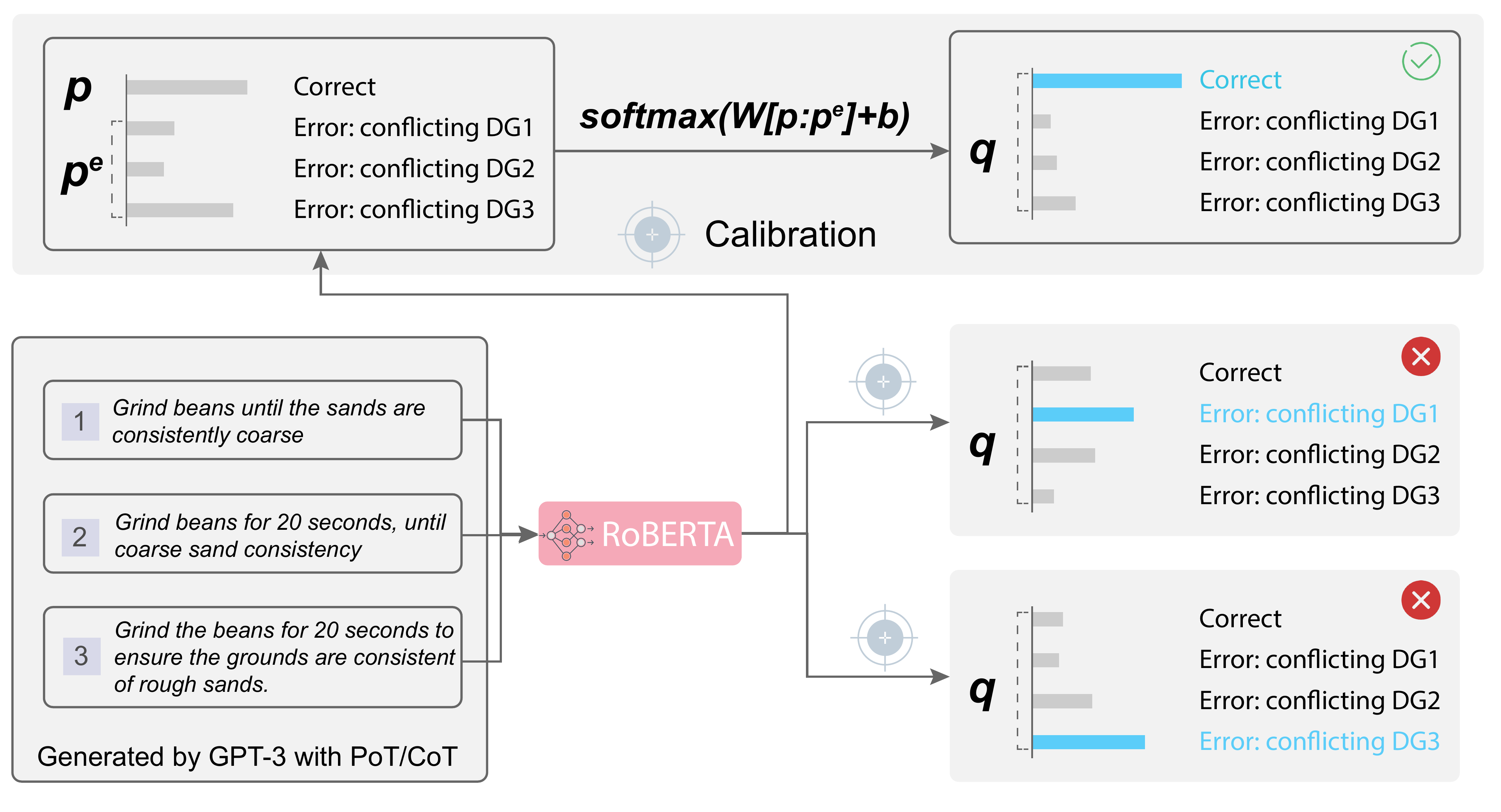}
    \caption{\revblue{Error-aware model calibration: \systemname~prompts GPT-3 to generate a set of candidate results, which are subsequently analyzed by a RoBERTa-based error classification model (depicted in pink block) to detect any violations of design guidelines. The predicted scores of errors are calibrated with the affine matrix. Scores are adjusted using an affine matrix to ensure that the final selection is the output with the highest probability of correctness. 
    }}
      \label{fig:calibration}

\end{figure*}
To align with~\ref{design_guideline:DG_1} and prioritize meaning preservation in text simplification~\cite{siddharthan2014survey, beigman2004text}, we propose an error-aware calibration method. Outputs from LLMs are often unstable and exhibit a bias toward certain answers due to the intrinsic bias of the LLMs and the influence of the prompt text, especially when applied to new tasks. Text simplification in AR has requirements that differ significantly from those of traditional NLP tasks, potentially exacerbating the impact of such intrinsic bias on LLM inference. For instance, LLM outputs may disproportionately reflect the influence of the last example in the prompt text, and within the context of our text simplification, the simplification techniques chosen may also be biased by the simplification techniques used in this last example~\cite{DBLP:conf/icml/ZhaoWFK021}. To mitigate these issues, our error-aware calibration mechanism adjusts the output probabilities by applying an affine matrix~\cite{brier1950verification}, which is learned from a set of annotated datasets. This transformation does not directly rely on the prompt text and can alleviate LLM bias~\cite{DBLP:conf/acl/YeD22, DBLP:conf/nips/YeD22, DBLP:conf/acl/ZhangGC21, si-etal-2022-examining}. Moreover, to strengthen LLM against common errors in AR, we enhance the annotated dataset with negative samples that violate~\ref{design_guideline:DG_1} and risk altering the original meaning.

 Shown in Figure~\ref{fig:calibration}, we use model calibration to stabilize language models in text generation \cite{DBLP:conf/acl/YeD22, DBLP:conf/nips/YeD22}. The affine transformation is defined as:
\begin{equation}
q = softmax(Wp + b),
\label{eq:calibration}
\end{equation}
where $p$ refers to the probability of the generated simplified text, $q$ is the calibrated probability, and $W$ and $b$ are learned parameters. We simplify this computation by treating $W$ as a diagonal matrix following~\cite{DBLP:conf/icml/ZhaoWFK021, DBLP:conf/icml/GuoPSW17}. The calibrated errors are identified from our open-ended exploration and expert interview. We then use the RoBERTA model to predict these errors by comparing the simplified text $T^*$ and the original text $T$ as, $p^e = {p^e_1, p^e_2,...p^e_m}=f(T, T^*),
$
where $p^e_i$ is the probability of an error and $m$ is the total number of errors.
\revblue{The errors include altering the meaning~\ref{design_guideline:DG_1}, producing text that is syntactically complex ~\ref{design_guideline:DG_2} and or too long~\ref{design_guideline:DG_3}.}
\revblue{Since access GPT-3's weights are not publicly accessible, we use RoBERTa instead to predict the error label $p^e$~\cite{DBLP:journals/corr/abs-1907-11692}.
}
Therefore, we modify Equation \ref{eq:calibration} by incorporating $p^e$,
\begin{equation}
q = softmax(W[p;p^e] + b),
\end{equation}
For each original text sample, we generate $n=5$ simplified text samples and calibrate them. The final output is determined by the calibrated probability.
\revblue{The parameter values of $W$ and $b$ are learned from a small set of manually crafted data samples\cite{DBLP:conf/nips/YeD22}. We first craft a set of gold-standard text simplification samples (64) $D=\{(T_1, Y_1, \hat{q}_1), (T_2, Y_2, \hat{q}_2 ),\\ \cdots ,(T_k, Y_k, \hat{q}_k)\}$ where $(T, Y, \cap{e})$ refers to the input text $T$, the simplified result $Y$, and whether the erroneous indicator $\hat{q}$. $\hat{q}$ indicates whether $Y$ is correctly simplified from $T$ and, if not, labels the error in $Y$. Since $W$ and $b$ have limited dimensions, we can learn the values of $W$ and $b$ through the gradient descent with the logistic loss function $|q\cap{e}|^2$. }
\begin{equation}
\begin{aligned}
    \mathcal{L}  & =-\hat{q}log(q) - (1-\hat{q})log(1-q) \\
      & = -\hat{q} log (softmax(Wp + b)) \\ &  - (1-\hat{q})log(1-softmax(Wp + b)),\\
  \end{aligned}
\end{equation}
where $\mathcal{L}$ is the loss function used to learn $W$ and $b$.

\subsection{Elaborative Simplification with Spatial Information}
\revblue{Following~\ref{design_guideline:DG_2}} and implementing elaborative simplification, we enrich the AR text by generating information on the spatial location of objects and object dimensions if they are presented in the original text.
\revblue{The object is detected and located with the Detic model, which runs on the backend server and provides the spatial information to LLM~\cite{zhou2022detecting}. 
As shown in the expert interview, many objects may exist in the working environment, while only a subset of them can be useful. To align with~\ref{design_guideline:DG_2}, we require LLMs to select the objects that are mentioned for the first time in the text.} Identified object locations are used to signify a spatial relationship to the user, adding a layer of contextual understanding that goes beyond identification.  For example, the text \textit{Then place the coffee mug with the dripper} can be elaborated with the coffee mug's detected position \textit{on your right} to form the result \textit{Then place the coffee mug on your right with the dripper}.
The spatial location is determined before the user clicks \textit{next step} and the elaborated content does not change during the execution of the step to avoid distracting the user. 
\revblue{The Detic model may incur prediction errors and mismatches in object location due to user movement and latency. For instance, the user's movement can alter the object's location relative to the user, and the Detic model's results may continue to indicate the location of the object before the user's movement. We mitigate this issue by predicting only the spatial relationships between the object and the user (e.g., \instructionquote{the object is to the right of the user}). Therefore, the minor errors and latency in the Detic model do not significantly impact the final result. Furthermore, in our open-ended exploration, we observed that during the step transition, users typically do not engage in significant movement, thereby reducing the likelihood of potential mismatches.}
When displayed text includes a numerical measurement and an object with comparable dimensions is identified in the AR environment, the system automatically substitutes the numerical value with a description of the detected object. 


\subsection{System Implementation Details}
We implement \systemname's functionality using OpenAI's GPT-3 APIs and build a text simplification server with Flask. We run the Detic model on the server and incorporate its object detection result into the prompt text for GPT-3. 
The interface is developed using the PTGCTL architecture~\cite{ptgctl}, with the HoloLens 2 component implemented in Unity.

\section{Evaluation}
\begin{table*}[ht]
\begin{tabular}{|r|p{1cm}|p{5.5cm}|p{5.5cm}|}
\hline
Task & \multicolumn{1}{|c|}{Step} & Original                                                                                                                                                                                                      & Simplified                                                                                                 \\ \hline
\multirow{2}{*}{Task 1.1} & 1                          & To create a coffee, first please carefully place the pour-over dripper over the coffee mug.                                                                                                                  & { Place dripper (on your left) on coffee mug.}                                            \\ \cline{2-4}
&                
7                          & Transfer the coffee grounds to the filter cone. Then place the coffee mug with the dripper on a digital scale and set it to zero.                                                                            & { Move grounds to filter cone. Set coffee mug with dripper on scale, zero it.}            \\ \hline
Task 1.2 & 2                          & Once the desk is clear, bring the power strip on the desk and connect the Charger to the power strip so the meeting attendants can use.                                                                           & Put power strip on desk, connect phone charger to it.                                                                        \\ \cline{2-4} 
& 5                          & Next, place cups of water and papers on each chair. Each person should have one cup of water and paper;                                                                                                           & Place water, paper onto desk in front of chairs.                                                                             \\ \hline

\end{tabular}
\caption{Four example system outputs in Study 1. The original, unmodified text (baseline) is in the third column; the last column shows the simplified condition with text output from~\systemname~ showing on the last column. }
\label{tab:study1-samples}
\end{table*}
The formative study elicited that text-based AR guidance often creates a high cognitive load and that following AR guidance can be challenging due to the HMD's small display, low readability, and user error. Although our system attempted to address these limitations by integrating text simplification into AR, the actual effects on users' cognitive load, performance, and sense of usability required further exploration. To better understand these effects, we conducted two empirical studies. The first (Study 1) focuses on the overall cognitive cost of our system and its effect on performance over unmodified text, and the second (Study 2) focuses on a comparison against other AR text simplification methods and what can be learned from them. Tasks in both studies are everyday tasks that could benefit from AR task guidance~\cite{quiroz2018emotion}. Both studies comprise within-subject designs. Although subtasks in Study 1 have a between-subject component, our primary focus and point of investigation is the text condition. We investigate the following research questions:
\begin{enumerate}
    \item In what ways does our proposed method impact cognitive load in AR (\ref{req:RQ_1})?
    \item In what ways does our proposed method affect task performance with text in AR (\ref{req:RQ_2})?
    \item How does our proposed method compare to other text simplification methods in AR (\ref{req:RQ_3})?
\end{enumerate}

\revblue{While the first study focuses on~\ref{req:RQ_1} and \ref{req:RQ_2}, the second explores {\ref{req:RQ_3}}. We pre-determined the study order so that half of the participants start with Study 1 and the other half with Study 2. Regardless of the order in which they engage the studies, participants are asked to review the study procedures and can only continue after giving their consent on the IRB-approved consent form.}

\subsection{Participants}
\revblue{Both studies involve 16 participants (average age 25, nine male and seven female). Half have previous experience using head-mounted AR and were recruited through electronic flyers and emails using snowball sampling.}

\revblue{\subsection{Study 1: The Effect of Text Simplification on Guidance Tasks}}
\revblue{We conducted an empirical study to evaluate the effect of textual simplification on users’ cognitive load, performance, and other subjective ratings. We select two common physical tasks that benefit from AR guidance and collect data from real users. To avoid the learning effect while keeping task difficulty levels similar, both subtasks are physical activities that are performed in the same room (See Figures~\ref{fig:study1.1-task} and~\ref{fig:study1.2-task}), have instructions of similar lengths, and do not require prior knowledge. }

\subsubsection{Experiment setup} We present the involved tasks and conditions in Study 1.

 \myheading{Task.}
The task contains two similar subtasks that have sequential instructions to guide users. In both subtasks, we display the AR text in a dark grey box to ensure visibility. We also adjust the font size to 9pt and have participants confirm that all text is legible. No single instruction is long enough to be cut off by the display. In terms of subtask assignment, we alternate the order of subtasks for each participant to balance the order effect.
%

\begin{figure*}[t]
    \centering
    \includegraphics[width=1\linewidth]{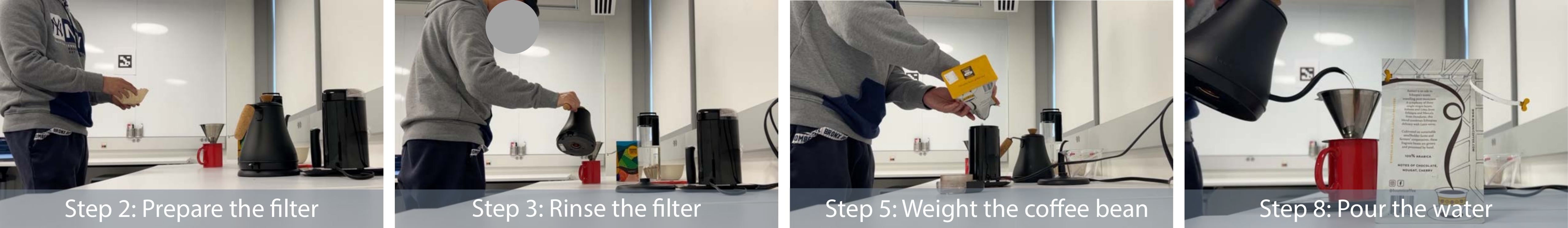}
    \caption{Task 1.1: Sample frames from user recording. \revblue{The task requires participants to make pour-over coffee based on a nine-step online tutorial. The frames were sampled from steps 2, 3, 5, and 8.} }
    \label{fig:study1.1-task}
\end{figure*}
\begin{itemize}
    \item \textit{Task 1.1: Pour-over Coffee.} This subtask contains nine step-by-step instructions that guide participants to make a pour-over coffee (Figure~\ref{fig:study1.1-task}). The instructions are taken from an online tutorial on \instructionquote{how to make pour-over coffee.}~\footnote{https://www.wikihow.com/Make-Pour-Over-Coffee} Participants need to read the text to complete the task.
    \item \textit{Task 1.2: Meeting room preparation.} This subtask requires that participants follow AR instructions to arrange objects in a meeting room based on a seven-step office menu (Figure \ref{fig:study1.2-task}). The instructions are digitized from an online manual.
\end{itemize}

\begin{figure*}[t]
    \centering
    \includegraphics[width=1\linewidth]{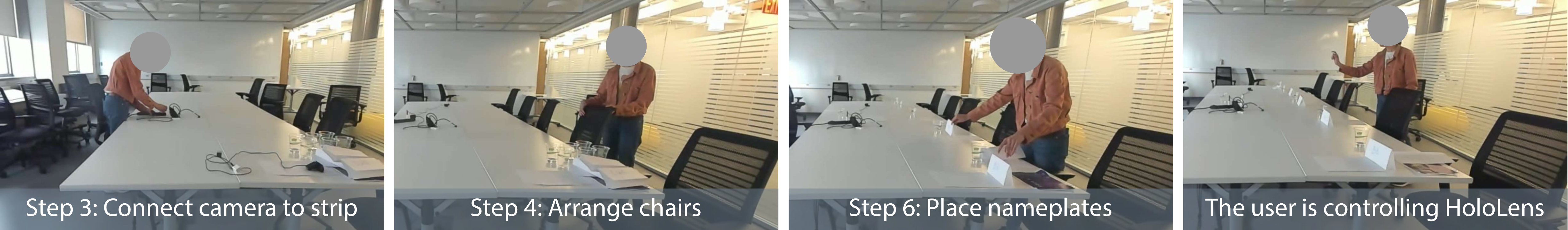}
    \caption{Task 1.2: Sample frames from user recording. The task requires the participants to arrange objects in a meeting room based on a seven-step office menu. The frames were sampled from steps 3, 4, and 6.   }
    \label{fig:study1.2-task}
\end{figure*}

\myheading{Conditions.}
\revblue{This study has two conditions: a baseline condition that uses the original imported text and a simplified condition using~\systemname. Each participant will perform one subtask (either Task 1.1 or 1.2) with the baseline condition and the other subtask with the simplified condition. We use a pre-generated table to alternate the order of all trials so that each participant will perform tasks in different orders under both conditions. In total, all conditions and subtasks are evaluated an equal number of times. Samples from the simplified and baseline condition can be found in Table~\ref{tab:study1-samples}.}  

\myheading{Apparatus.}
\revblue{Participants wear a Microsoft HoloLens 2 and use hand gestures and voice commands to interact with the AR menu. These interactions are native to HoloLens 2, and the AR interactions comprise standard button tapping, translating, and spatial movement. Video and audio recording devices are set up to collect participants’ feedback and qualitative data.}

\subsubsection{Procedure}
 
\revblue{The experimenters welcome the participants in a physical room; the physical tools necessary for task performance (e.g., coffee machine and ingredients) are present. To maintain ethical standards and comply with the IRB guidelines, each participant is given an informed consent form before the evaluation. Upon signing, each participant is paid an hourly rate of $\$20$ and is fitted with the Microsoft HoloLens 2 headset. An ill-fitting HoloLens 2 can be detrimental to the AR experience, causing blurry text. A series of initial calibrations are performed to ensure interface functionalities~\ref{fig:demo-teaser}.}

\revblue{
After all participants successfully interact with the AR interface, including its menus and buttons, using hand gestures, and indicate they can see the AR text clearly on the HMD, experimenters then explain the two subtasks and ask participants to practice thinking aloud. Meanwhile, video and audio recordings were set up before the trial began. Participants begin the study by air-tapping the AR button marked \instructionquote{Start} at the center of the HMD’s screen.}

\revblue{
Once the task starts, step-by-step text instructions are automatically displayed in AR. Participants are not informed which condition they are using and are asked to think aloud while we observe and record the trials. Any anomalies or potential safety issues are continuously monitored by the experimenter. At the end of each subtask, we collect the participant’s subjective ratings on text readability, comprehensibility, guidance performance, trust, and cognitive load using a NASA TLX form. A semi-structured interview is conducted to better understand their experience.}

\begin{figure*}
\centering
  \includegraphics[width=\textwidth]{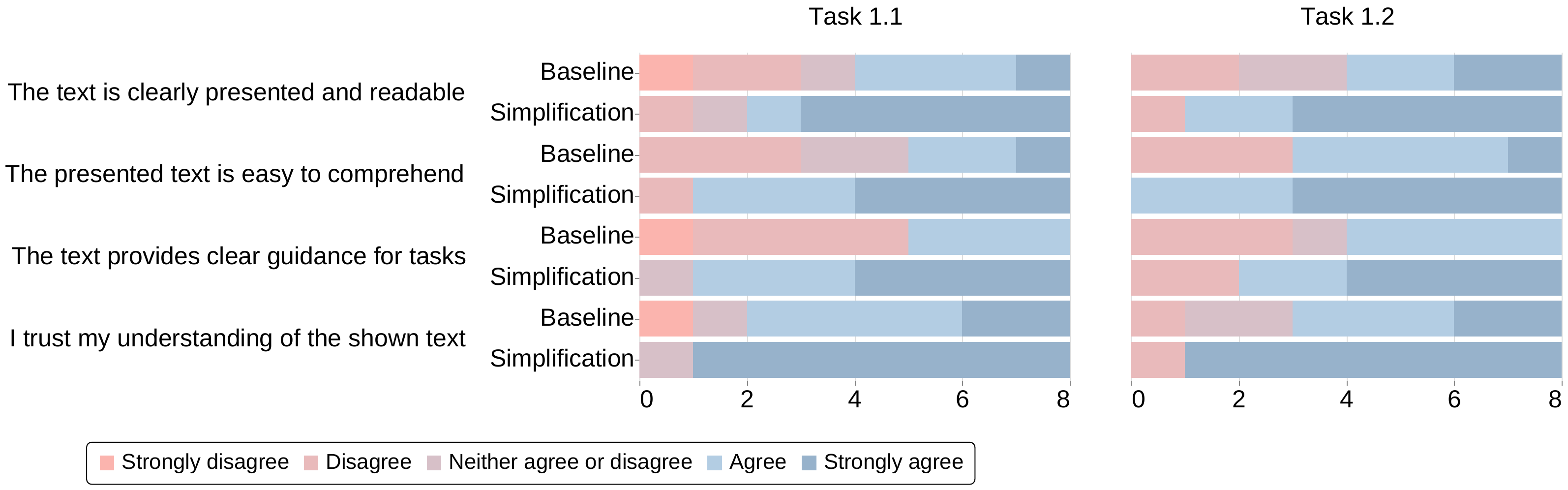}
\caption{Study 1 results on a five-point Likert scale. Ratings are collected on a scale from ``strongly disagree'' to ``strongly agree'' in response to four questions assessing the readability, ease of comprehension, guidance, and trust in both simplified and baseline text versions. The horizontal bar graphs above visually  represent the distribution of these ratings. The distribution reveals more positive responses for the simplified text across all questions and tasks.}
\label{fig:study1-uq}
\end{figure*}

\begin{figure*}
\centering
  \includegraphics[width=0.8\textwidth]{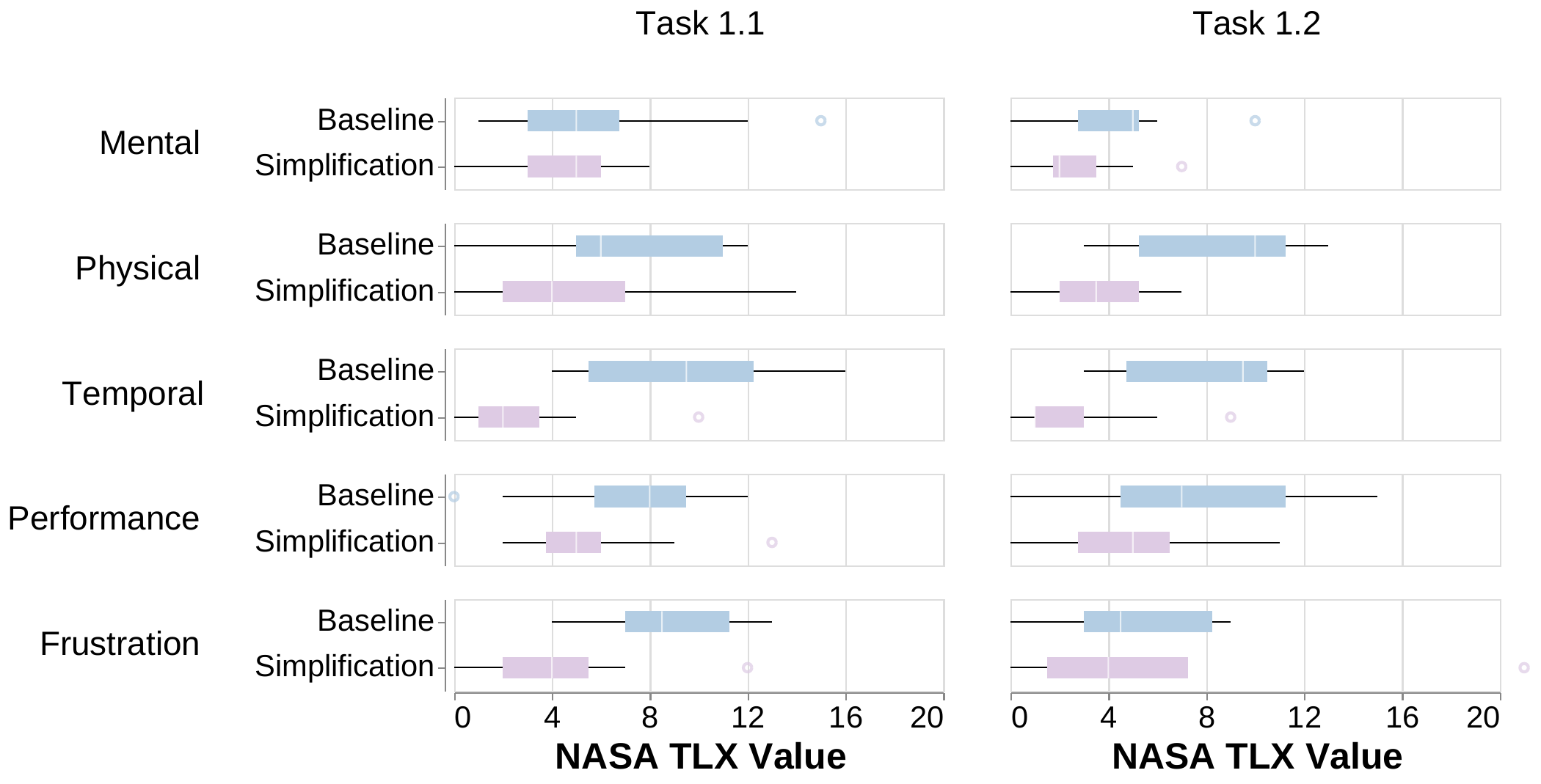}
\caption{Study 1 results on NASA Task Load Index (TLX) values. \revblue{The y-axis represents the different aspects of the NASA TLX, while the x-axis shows the TLX values. The simplified text significantly surpasses the baseline in reducing temporal demands and in enhancing performance and reducing frustration, demonstrating its advantages for overall task load.}}
\label{fig:study1-tlx}
\end{figure*}

\subsubsection{Data collection} 
\revblue{We collect quantitative data to measure performance. We specifically record the number of errors and the number of steps participants recall (i.e., memorability); in addition, we explore self-evaluated performance via subjective ratings. The experimenter counts the number of errors during participants' trials. Memorability is measured because one major challenge in AR guidance is that users only recall limited AR information during physical tasks; remembering steps reduces the need to split attention between AR and the task. Subjective ratings are inspired by the System Usability Scale (SUS), and we collect five-point Likert ratings on AR text readability, comprehensibility, guidance, and trust. We explain that trust reflects how confident the user is with their task performance.}
\begin{figure*}
\begin{tabular}{cc}
  \includegraphics[width=70mm]{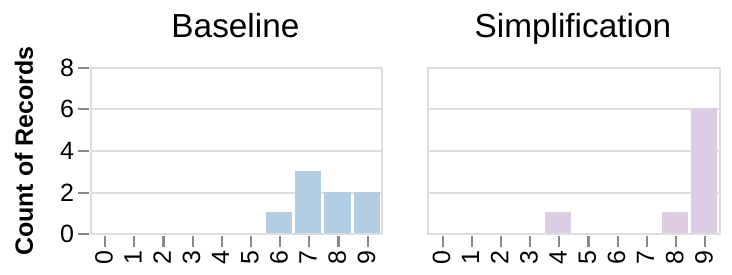}  
&
 \includegraphics[width=70mm]{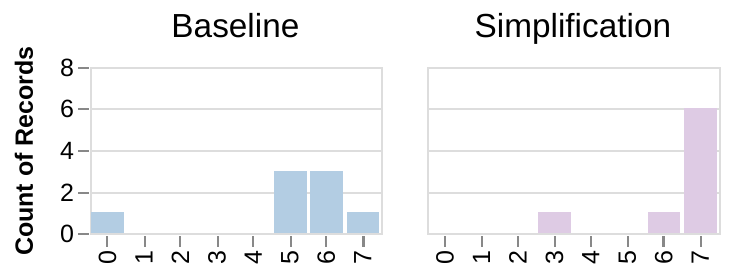} \\
 (a) \revblue{Number of recalled steps in Task} 1.1  & 
(b) \revblue{Number of recalled steps in Task} Task 1.2 \\
\end{tabular}
\caption{\revblue{Study 1 results on the number of errors participants made while performing Tasks 1.1 and 1.2. The x-axis indicates the number of steps successfully recalled, while the y-axis shows the count of participants who recall that number of steps. 
}}
\label{fig:study1-memory}
\end{figure*}

\begin{figure*}

\includegraphics[width=0.8\textwidth]{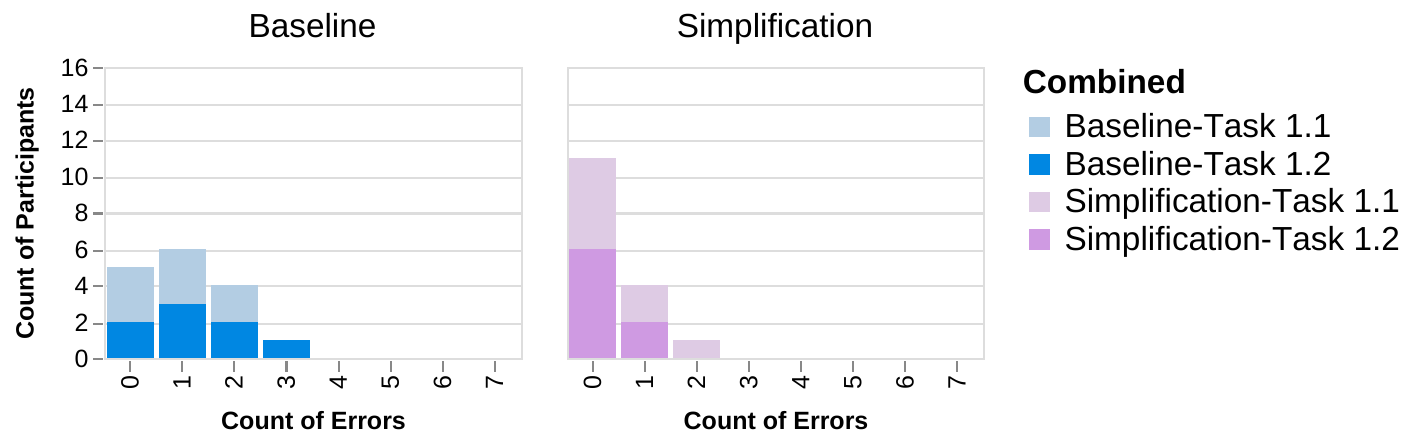} 
\caption{\revblue{Study 1 results on the number of errors made in Tasks 1.1 and 1.2. The bar graph compares the error count between the baseline and simplified conditions, with the x-axis recording the number of errors and the y-axis depicting the number of participants who make those errors. The data illustrates that participants commit fewer errors when following the simplified text.}}
\label{fig:study1-error}
\end{figure*}
\revblue{Cognitive load is a primary user performance limitation in AR guidance tasks, and we use a NASA TLX 8(a)(b) form to measure it. Raw TLX scores are used and summative results are analyzed based on Hart's recommendations~\cite{hart2006nasa}.}

\revblue{Experimenters also collect qualitative data via video and audio recordings of the study. Interview notes, think-aloud notes, and observations are also collected for later analysis. The sampled frames for the study can be found in Figures~\ref{fig:study1.1-task} and~\ref{fig:study1.2-task}. }



\subsection{Study 1: Results and Discussion} 
\subsubsection{Quantitative results}
\revblue{Using Mann-Whitney's U test, we assess differences among TLX scores, recall, and error data and use the Friedman test to assess differences among subjective ratings. These tests were chosen because the data are non-parametric. The TLX analysis shows that the simplified condition significantly reduces the overall cognitive load for both subtasks~($U = 52,~z = 2.84,~p < 0.01$). Further evaluation of recall and error found no significant improvement in recall for the simplified condition over the baseline~($U=85,~z = 1.63,~p = 0.10$), but showed the simplified condition significantly reduced the number of errors counted by users over the baseline~($U=72.5,~z = -2.09,~p=0.024$), see Figure~\ref{fig:study1-memory}. Test on subjective ratings indicated significant differences in all categories: readability ($\chi^2 = 10.71,~p = 0.013$), comprehensibility ($\chi^2 = 15.00,~p = 0.002$), guide~($\chi^2 = 10.71,~p = 0.013$), and trust~($\chi^2 = 18.00,~p = 0.001$).} 

\begin{figure*}[t]
    \centering
    \includegraphics[width=0.85\linewidth]{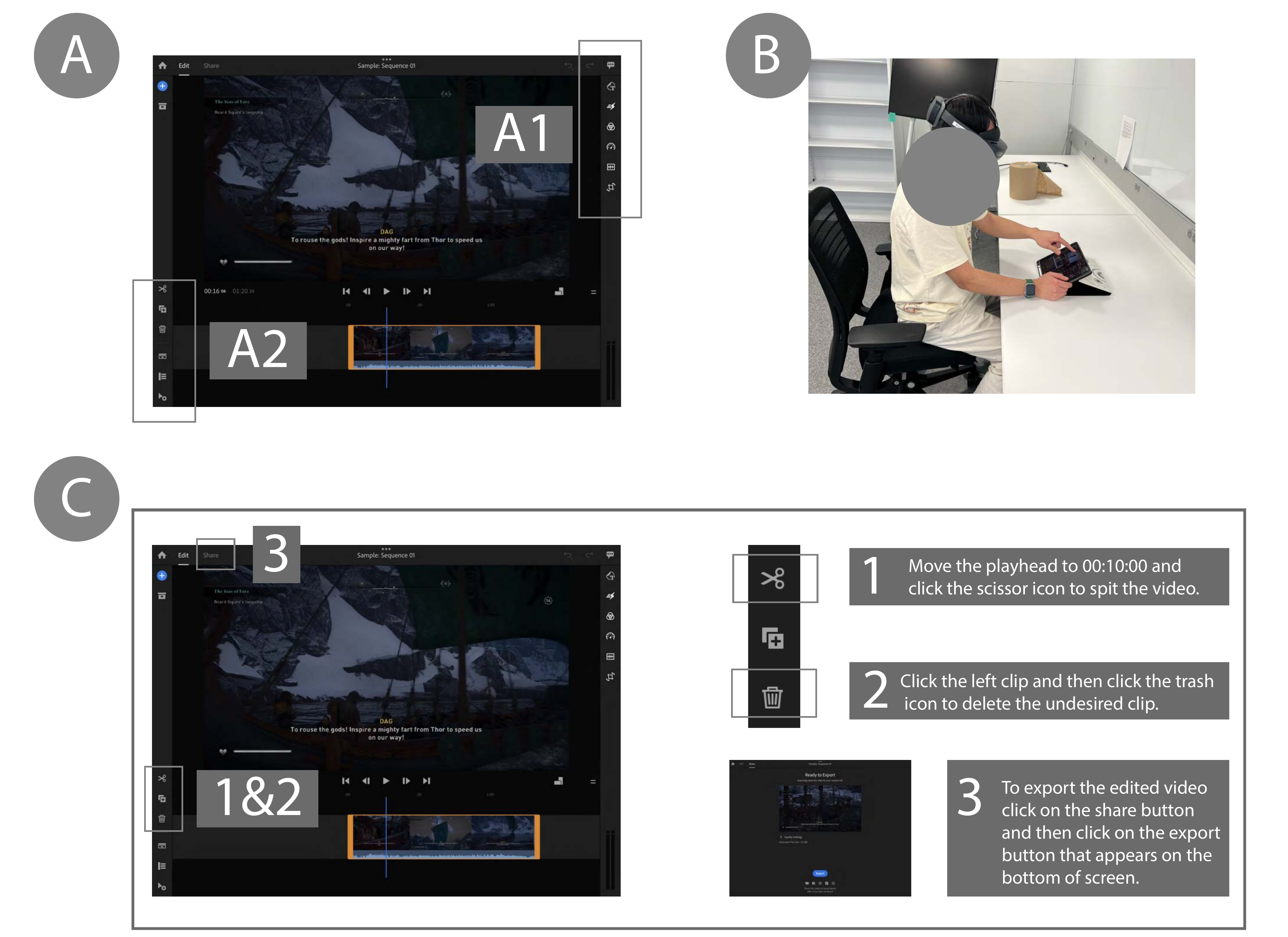}
    \caption{(A) Adobe Premiere Rush Interface: The interface showcases a video player positioned at the center of the screen, accompanied by a timeline below. The top-right section (A1) features buttons for graphics, effects, color, speed, audio, and transform functionalities. The bottom-left section (A2) contains buttons for editing tools and the project panel. (B) User Record: This section captures the user's interactions and activities during the Study 2 session. (C) Example Task Description: An illustration of a sample task description used in the study, providing users with instructions for completing a specific editing task.}
    \label{fig:study2-task}
\end{figure*}

\subsubsection{Qualitative results} 
We coded the transcribed video and audio data along with notes from thinking aloud and observation. Codes sharing similarities were then grouped into themes to summarize analogous findings.

\myheading{Spatial information can assist users.} \revblue{Participants acknowledged the benefits of spatial information in reducing cognitive load. Elaboration on objects’ spatial location eliminates the need to search for them, reducing user effort. P11 reported feeling nervous when presented with multiple objects and new mentions of objects in the AR interface. P11 mentioned, \participantquote{Sometimes it is overwhelming to face many objects, and the location word (on your left) helps (you) find the object.} The reduced effort and pressure were also confirmed by P1, who stated, \participantquote{Even though it won't save much time, the elaboration on the object eases my (sense of) pressure.}  }

\myheading{Text length and structural complexity affect participants’ performance.}
\revblue{Most participants report that shorter text is beneficial. Some participants reported that shorter text takes less time to process (P2, P5, P7) and felt it was \participantquote{easier to understand}  each step during a subtask when the text was shorter (P6-7). This is reflected in the TLX scores, as simplified conditions yielded better cognitive load scores than the original texts. }
\revblue{Participants further report that using shorter sentences leads to better comprehension and confidence (P2, P10-12, P14). Multiple participants pointed out that they naturally \participantquote{skim} text in AR, and stated that complex sentence structures lead to skipping important information and misunderstandings. More than half of the participants further stated that the simplified text improved their trust. When asked to explain their reasons for skimming text, screen resolution, screen size, and the urgency of completing physical tasks (impatience) while wearing a headset were identified. These observations reflect what experts from the formative study indicate. }

\myheading{Simplified text improves task guidance.}
\revblue{
Participants respond positively to breaking longer sentences into shorter ones (syntactic simplification). \systemname~ divides long sentences into shorter ones by adding verbs (elaborative simplification). P5 said \participantquote{Shorter sentences with clear actionable directions make it easy to know what to do,} while P6 said, \participantquote{It is more convenient to follow smaller step instructions.} The participants’ positive feedback reflects the benefits introduced by the design guidelines proposed earlier. }

\subsubsection{Discussion}
\revblue{
Our results verify that \systemname~ significantly improves cognitive load (\ref{req:RQ_2}) and reduces task performance errors, generating significantly higher subjective ratings (\ref{req:RQ_1}). Shortened sentences and syntactic simplification contributed to decreased cognitive load, as participants indicate that the simplified text is easier to read. Additionally, shorter sentences enable participants to quickly skim the text to grasp core concepts, which may also play a part in reducing cognitive load. As we mentioned earlier, reducing cognitive load could help to improve the usability of AR guidance and have a positive effect on users’ safety. }

\revblue{\systemname~ significantly improves subjective ratings on all four metrics. However, the system yields no significant change in memorability. Both the baseline and simplified conditions reached a fairly high recall count. A possible explanation for this is the fact that all tasks are physical tasks, and participants may rely on their performance more than the text for recall. 
Yet the simplified text resulted in fewer user errors than the unmodified text, suggesting that the system successfully retains critical information for tasks. Overall, participants felt better guided by the simplified instructions and more confident (i.e., trust), signaling a positive effect on their overall performances. }


\begin{figure*}[t]
    \centering
    \includegraphics[width=0.95\linewidth]{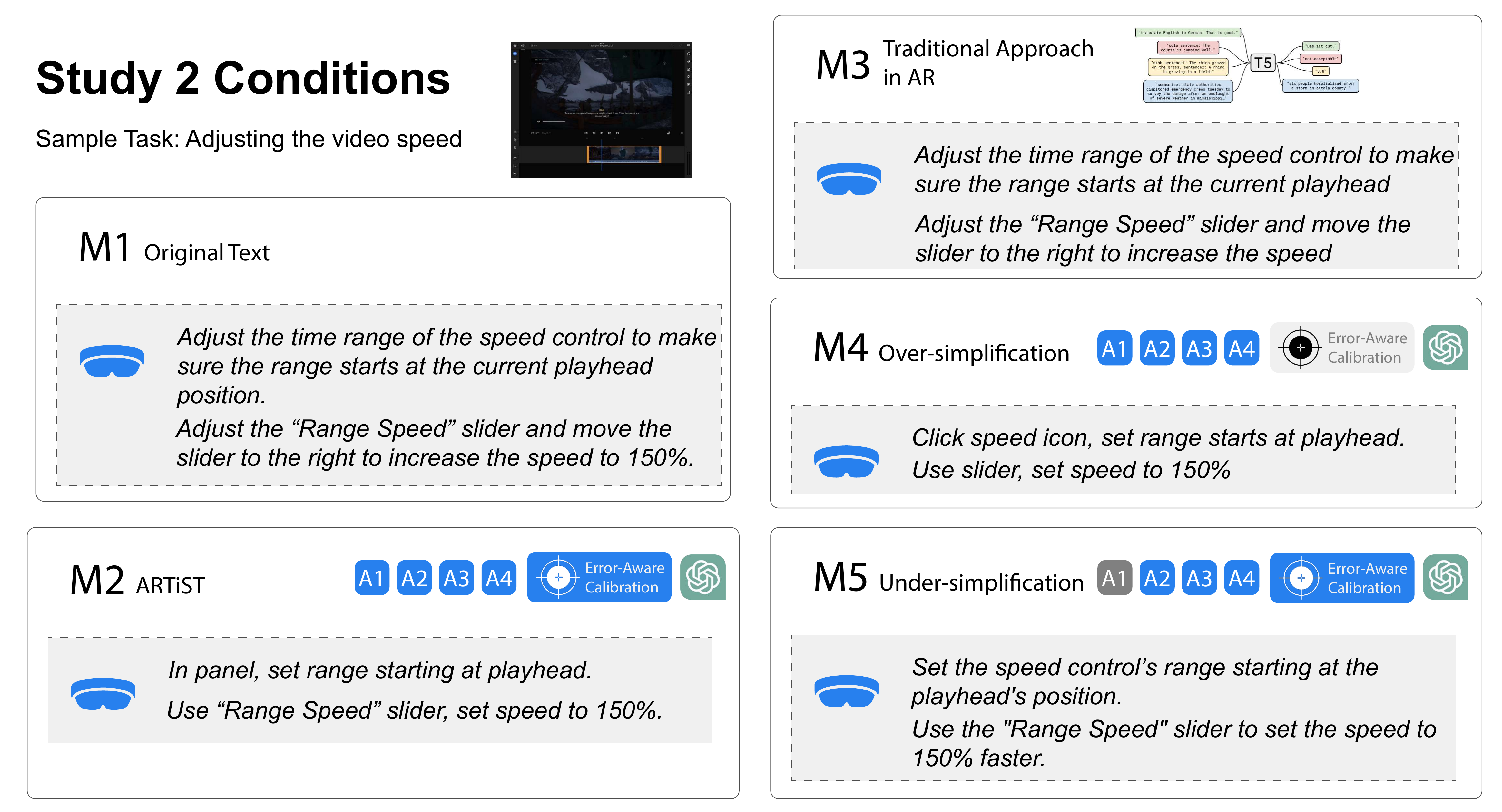}
    \caption{All five conditions for Study 2. M1 is the original text. M2 is the~\systemname~condition. M3 is the state-of-the-art T-5 model applied to AR. M4 is \systemname~ without engaging error-aware calibration (over-simplification). M5 is \systemname~ without engaging content reduction(under-simplification). The text in the method's grey box represents the text after simplification. A1, A2, A3, A4, and error-aware calibration legends denote whether any of these components are used for the condition.  
    }
    \label{fig:study2-conditions}
\end{figure*}

\subsection{Study 2: Comparing Text Simplification Methods} 
\label{sec:study2}\revblue{In the previous study, we evaluated \systemname~ against unmodified AR text. The goal of this study is to further understand how \systemname's process compares with other methods for text simplification. However, almost all currently used methods are not tailored for AR. As such, we selectively integrated these methods into the AR context while keeping their traditional functionalities. This study is a within-subject study that includes five different methods with one task. The study recruited the same set of participants ($N=16$). In addition to the HoloLens 2 used in Study 1, this study makes use of an iPad as an additional apparatus for task performance.}

\subsubsection{Experiment setup} We present the tasks and conditions of Study 2 in this section.

\myheading{Task.} \revblue{The task is designed to have participants wear a HoloLens 2 while also using an iPad. AR instructions for video editing are displayed on the HoloLens 2. To minimize the learning effect and for repeated trials, we employ subtasks with similar interactions and difficulty levels but different content. We adopt a series of video editing jobs from Adobe Premiere Rush's official tutorial to AR\footnote{Adobe Premiere Rush. \url{https://helpx.adobe.com/premiere-rush/tutorials.html}} to test each method. They involve interaction primitives such as selection, pan, and translation. The task contains five subtasks, including video clipping (S1), speed control (S2), graphics overlay (S3), video filter application (S4), and aspect ratio adjustment (S5). These subtasks are chosen because they have similar interaction difficulty but require diverse types of interactions (e.g., tap, drag, and pinch). Each subtask has three steps and takes about three minutes to complete based on our preliminary testing. Adobe Premiere is installed on the iPad participants use to perform the video editing. An example can be found in Figure \ref{fig:study2-task}.}

\myheading{Conditions.} \revblue{To understand how our system differs from other simplification methods and investigate the implications for user performance, we explore five conditions. Beyond making a comparison to the unmodified text (i.e., baseline), we also compare against the state-of-the-art T-5 text simplification model~\cite{DBLP:journals/jmlr/RaffelSRLNMZLL20}. Further, because our formative study revealed that sentence complexity and grammar structure (text length) have a foremost effect on text reading in AR (which was also indicated by experts in the formative study), we also explore an over-simplification and an under-simplification condition in this study. These two simplification methods represent different levels of length modification relative to the original sentence. Over-simplification is achieved by removing the error-aware calibration step for maximum simplification at the cost of factual information. Under-simplification is achieved by removing the content reduction technique in \systemname. We describe the five different conditions below:}

\begin{itemize}
    \item M1: Original text
    \item M2: \systemname's approach;
    \item M3: Traditional state-of-the-art text simplification with T-5 ~\cite{DBLP:journals/jmlr/RaffelSRLNMZLL20} fine-tuned on WikiAuto dataset \cite{DBLP:conf/acl/JiangMLZX20}
    \item M4: Over-simplification without error correction (i.e., does not force factuality, see~\ref{sec:dg} for details)
    \item M5: Under-simplification without content reduction~(\textbf{A1})
\end{itemize}

We use a pre-generated table to balance the learning and ordering effect for the five conditions across the five subtasks. \revblue{We assign one condition to one of the subtasks to form pairs, and each pair includes three steps (i.e., three trials). For example, the pair M2-S1 stands for the M2 condition used in subtask 1. The pre-generated table ensures that each participant performs these pairs in a unique order. Each participant performs five condition-subtask pairs or 15 trials, for a total of 240 trials. Overall, all subtasks and conditions are evaluated an equal number of times.}



\subsubsection{Procedure}
\revblue{In the beginning of this study, participants are asked to wear the HoloLens while sitting and holding an iPad. Experimenters explain that their task entails editing videos that are displayed on the iPad. The videos are 30-second clips of stock footage, and participants are told they will use the onboard video editing tool on the iPad to seek, crop, change filter, and change the aspect ratio. After a short warm-up period to familiarize participants with iPad functionality and fit the HoloLens, we confirmed that participants can read the AR text clearly, similar to what we did in Study 1. 
During task performance, condition-specific AR text is displayed to the participants; they are asked to follow the text to perform the editing task. Experimenters count the number of errors made during the trials, and participants are asked to think aloud as they engage in their tasks. At the end of each condition, the experimenter collects recall data, subjective ratings, and TLX scores. A semi-structured interview is conducted at the end of the study to understand the participants’ overall impression of each of the five conditions. We collect both quantitative and qualitative data in a similar way to Study 1.}

\subsection{Study 2: Results and Discussion} 
\subsubsection{Quantitative results}
\revblue{For non-parametric data, we used the independent-sample Kruskal-Wallis' test with repeated measures for performance metrics (error, memory recall, and subjective rating) and for cognitive load \revblue{with Dunn's Test as post-hoc analysis with Bonferroni correction for multiple tests. For error analysis, we found an overall significant effect ($H(4) = 17.189, p = 0.001$), with post-hoc analysis showing that M2 ($p=0.014$), M3 ($ p =0.014$), M4 ($p=0.014$), and M5 ($p=0.014$) reduced errors significantly compared to the baseline. We found that there is an overall difference across the conditions ($H(4) = 12.572, p = 0.014$) in terms of participants’ ability to recall the instruction steps. Post-hoc analysis revealed a significant difference between the original text M1 and  \systemname~($p = 0.025$). No differences are found between the original text M1 and M3 ($p = 0.179$), M4 ($p = 0.319$), and M5 ($p > 1.00$). No differences are found between the four simplified conditions (M1-4). As for TLX scores, we found that M2 significantly reduced overall cognitive load compared to M1 ($p = 0.043$): for a detailed breakdown refer to Figure~\ref{fig:study2-tlx-raw}. There are no significant differences among the five conditions in readability ($H(4) = 0.934, p = 0.934$), comprehensibility ($H(4) = 0.389, p = 0.983$), guidance ($H(4) = 2.444, p = 0.655$) and trust ($H(4) = 1.530, p = 0.821$); See Figure~\ref{fig:study2-uq} for details.} }

\subsubsection{Qualitative results} We identify a series of qualitative findings based on the quantitative metrics and the coded recordings.

\begin{figure*}
\centering
\includegraphics[width=\linewidth]{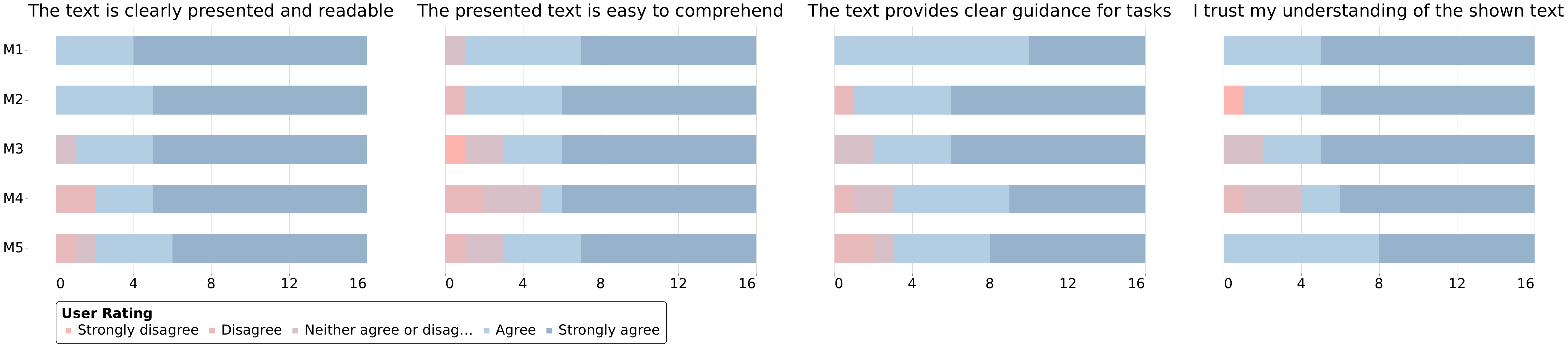}


\caption{\revblue{Study 2 results on the subjective Likert scale. Ratings were collected on a scale from ``strongly disagree'' to ``strongly agree'' in response to four questions assessing the readability, ease of comprehension, guidance, and trust in both simplified and baseline text versions. The horizontal bar graphs represent the distribution of these ratings, and the results for the four questions are laid out horizontally.}}
\label{fig:study2-uq}
\end{figure*}

\begin{figure*}
\centering
\begin{tabular}{ccc}
  \includegraphics[width=65mm]{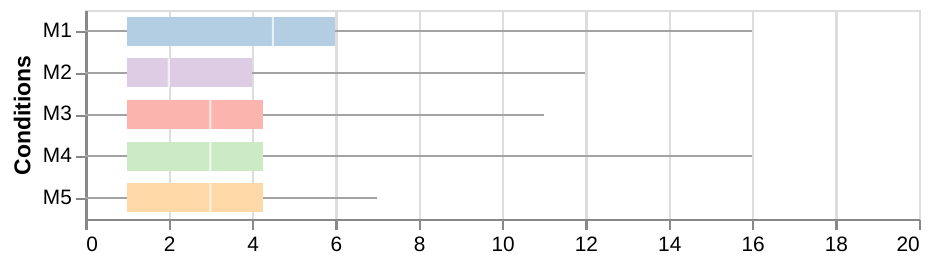} &   \includegraphics[width=45mm]{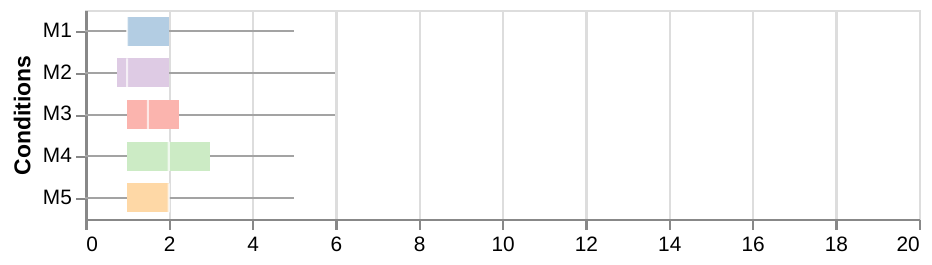} &  \includegraphics[width=45mm]{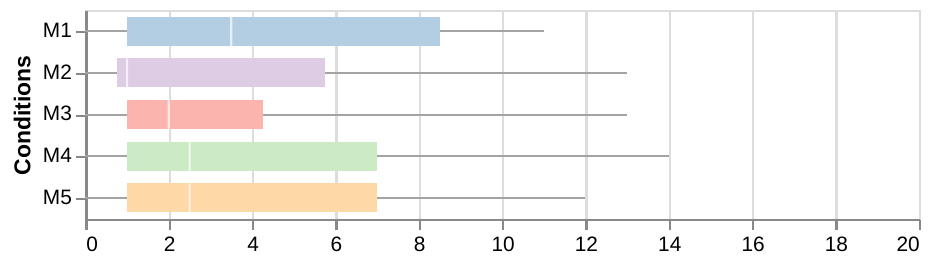} \\
(a) NASA TLX on mental demand & (b) NASA TLX on physical demand & (c) NASA TLX on temporal demand\\[6pt]
\\[6pt]
\includegraphics[width=65mm]{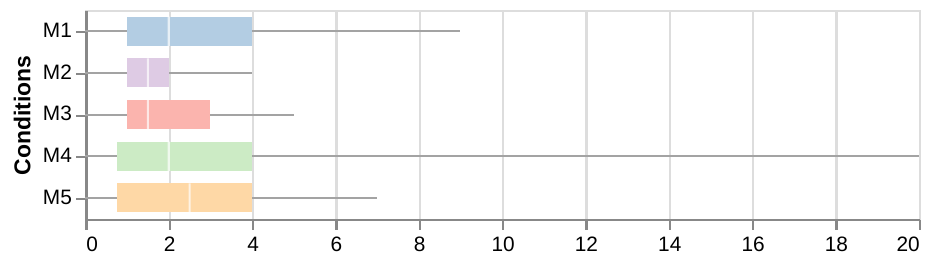} & \includegraphics[width=45mm]{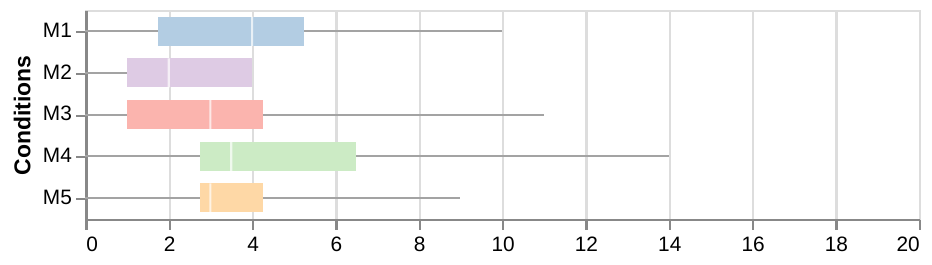} &   \includegraphics[width=45mm]{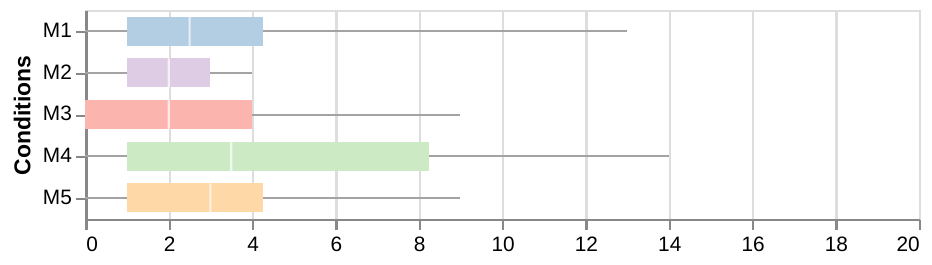} \\
(d) NASA TLX on performance  & (c) NASA TLX on effort & (d) NASA TLX on frustration \\[6pt]
\end{tabular}
\caption{\revblue{Study 2 results on NASA Task Load Index (TLX) values. The y-axis represents the different aspects of the NASA TLX, while the x-axis shows the TLX values. The results indicate that condition M2 significantly outperforms other conditions in terms of the user’s effort.}
}
\label{fig:study2-tlx-raw}
\end{figure*}

\begin{figure*}
\centering
\begin{tabular}{cccc}
  \includegraphics[width=85mm]{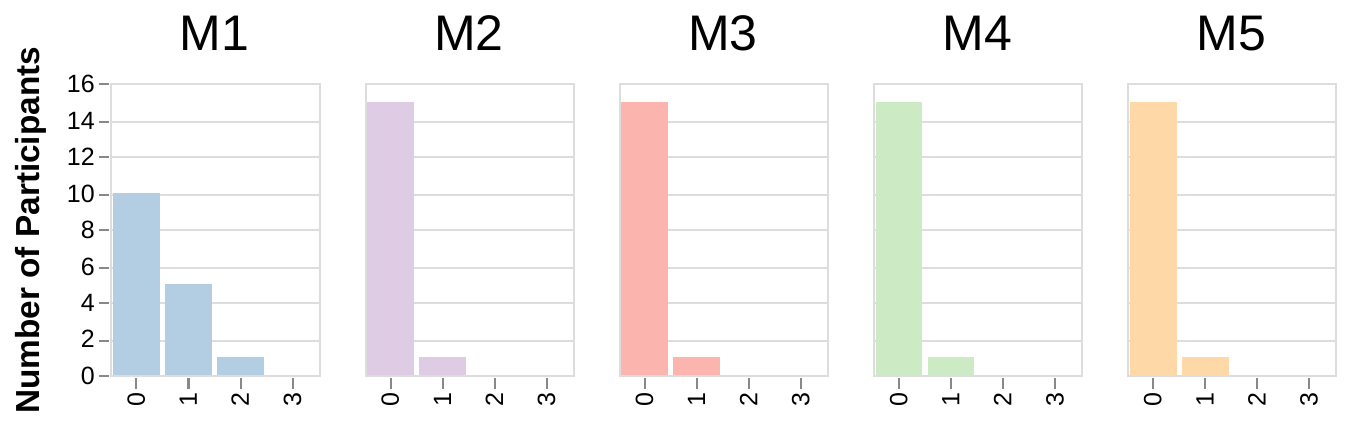} &   \includegraphics[width=70mm]{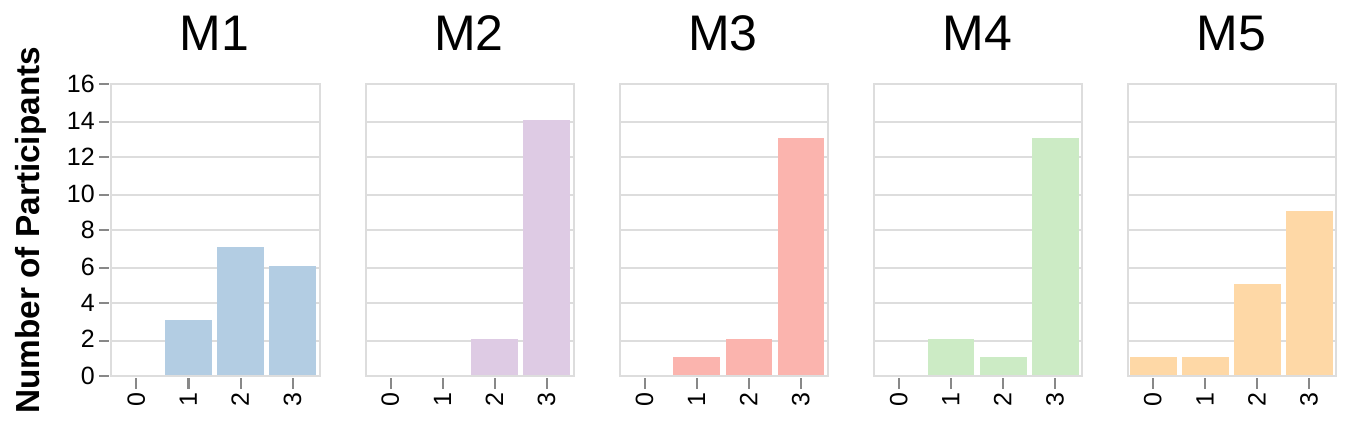}  \\
 (a) Number of errors made in the tasks  & (b) Number of the recalled step after the task \\[6pt]

\end{tabular}
\caption{
\revblue{Study 2 results for error count and memorability evaluation. 
In panel (a), the x-axis represents the number of errors made by participants, while the y-axis shows the count of participants corresponding to each error count. The results indicate that methods M2, M3, M4, and M5 are effective at reducing the number of errors, underscoring the advantages of text simplification in enhancing task success. In panel (b), the x-axis displays the number of task steps correctly recalled after the task, and the y-axis shows the number of participants. This panel demonstrates the impact of the simplification process on the participants' abilities to recall information, with conditions M2, M3, and M4 showing an improved number of steps recalled over M1.}}
\label{fig:study2-errro-memory}
\end{figure*}

\myheading{The level of simplification has a mixed effect on error rates and subjective ratings.}
While participants reported that they could understand any of the simplified texts (M2-5) better than the unmodified text (M1), we noticed that there is no uniform effect on the level of simplification relative to task errors (P11-12, P3, P5). \revblue{While several participants experienced increased errors due to over-simplified text omitting important information (P3), others made mistakes due to verbose text that was not simplified enough (P11-12, P5), causing them to overlook important information. This also aligns with the fact that the M1 condition yielded the most user errors. One of the behaviors observed is that the longer the sentences, the less patience a participant appears to have and the faster they skim. Often these behaviors lead to missing details while carrying out the task, such as when P11 and P12 adjusted the wrong button during task performance.}
Similarly, both over- and under-simplification methods affected participants' sense of readability and memorability in different ways. When asked about their experience with the M5 condition, P6 reported that \participantquote{the simplified one uses the more understandable words,} but P10 mentioned that the texts are \participantquote{not simplified enough and can be thrown away.}  P7 also reported that M5 increased the number of previous steps they could recall, as M5 presents \participantquote{clear and memorable instructions.}

\myheading{The effect of different text simplification methods on cognitive load.}
Participants reported during the interview that simplified texts (M2-5) have in lower cognitive load. While reading unmodified text became tedious during task guidance (P2, P9-10, P14), the simplified text could be less so (P2). However, participants reported that over-simplified text (M4) increases cognitive load, as important information is often removed resulting in extra processing time needed (P1, P10).  (P1, P10). Moreover, participants indicate that they believe they perform better with the \systemname~condition: As P7 mentioned, \participantquote{I am pretty sure I successfully completed the steps,} while P11 said, \participantquote{I feel it increased my performance.}

\subsubsection{Discussion}
\revblue{In exploring \ref{req:RQ_3}, we found \systemname~ impacted TLX ratings as it was the only condition that significantly reduced cognitive load for participants. Figure~\ref{fig:study2-tlx-raw} shows that TLX variance is much lower for performance with~\systemname, which is in line with our observation that most participants show stable performance with the~\systemname~condition. }

\revblue{All four simplified text conditions  (M2-5) significantly reduce error rate, but do not necessarily increase recall. This finding reflects our \ref{design_guideline:DG_3}, which addresses the importance of text length in AR. Regardless of the level of simplification, all four conditions shortened the text in some way. The results indicate that only \systemname~significantly improved recall while reducing error rate. This indicates that \systemname~helps users to improve performance in short-span tasks like video editing. In addition, the current state-of-the-art text simplification (M3) does not reduce high cognitive load nor improve memory for AR readers, while \systemname~improved on both. This suggests that the direct application of text simplification to AR might not be optimal and is in line with the results from the formative study. }

\revblue{\systemname~is also the only condition that significantly reduced the TLX scores (\ref{req:RQ_2}). Sentence length and structure may play an important part in reduced cognitive load as participants noted reduced processing time and more ready comprehension. This reduction addresses the concerns (high cognitive load) brought up by experts during our formative study in Sec.~\ref{sec:expert}. }

\revblue{Finally, both over- and under-simplification conditions received mixed feedback from participants. This could be linked to their personal reading habits when wearing an HMD. We observed that participants who comment positively on the over-simplified condition are typically impatient readers when they have the HoloLens 2 on. Others, however, complained that the over-simplified condition does not provide enough detail or is missing critical information, creating obstacles to task completion. Our qualitative results showed that participants who took extra effort going after missing details scored higher in their TLX ratings. These findings reflect the results from the formative study that both text length and meaning preservation are important.}



\section{Discussion}
\subsection{The Scope of \systemname}
While our evaluation showed improved performance and reduced cognitive load for participants using  \systemname, the scope in which our system works could be affected by task types (stationary vs mobile), task difficulty, and users' expertise. \systemname~ shows promising results for relatively stationary tasks, such as making coffee or preparing meeting rooms, where users can pause between subtasks to read instructions. As a result, we envision that tasks related to medical diagnostics and work productivity can benefit from text simplification. Text-based guidance may perform poorly for tasks where users are in constant motion (e.g., riding a bike or running outdoors), as reading could divert attention and pose safety risks. 

\revblue{Task difficulty could further affect the applicability of text-based instruction. The pure text might not be the best medium to guide highly complex tasks such as surgery or repairing mechanical devices, which may benefit from multimedia guidance comprised of visual or visual-auditory media alongside text. However, the concepts of textual simplification may extend to multimedia guidance and future work could be done to explore the simplification of additional variables (e.g., layout, typography, and visual elements) in the context of more complicated AR tasks. }

\revblue{Finally, the user's own level of expertise also affects the experience of using \systemname. Experienced users tend to skip instructions and may desire very minimal guidance is a known behavior. A beginner, on the other hand, might benefit from more elaborative text guidance. This creates an opportunity to personalize text simplification for users with different levels of expertise and different guidance preferences.}

\subsection{Implications of AR Text Simplification} 
\myheading{Spatial information benefits AR users.}
 A command like \instructionquote{the cup on your right,} facilitates object identification. Participants give positive feedback about spatial elaboration as this method helps to contextualize content in the original text. Future work could explore how we can further contextualize text via AR animation or AR sound. 

\myheading{Object detection error rarely impairs task performance in AR context} 
\revblue{We use object detection (i.e., Detic~\cite{zhou2022detecting}) to gather the spatial information. We acknowledge that the Detic model may not always be stable, and its errors could potentially impair task performance. In our study, we did not observe any significant issues with the Detic model. There are two possible failures: false positives and false negatives. We observed no false positives but some false negatives, where objects were not detected by Detic. 
Interviewing users who encountered false negatives found that these false negatives were unnoticeable by participants because text was not always simplified; this did not lead to users generating more errors. Although no false positives were encountered, we hypothesized situations where incorrect object locations might be presented. Participants responded that such issues would not significantly impair their performance, as they felt they would identify the error when the object was not found in the indicated location.}

\myheading{Integrating empirical findings from AR into LLMs may open up new opportunities for human-AI collaboration.}
Emerging human-AI collaboration tasks often require knowledge about users' needs. This knowledge often remains undeveloped until empirical studies are conducted~\cite{DBLP:conf/chi/ZhengWSMLM23, DBLP:conf/chi/WangCMFSSW20, 10.1145/3544548.3581469, DBLP:conf/cvpr/WuLS22}. The traditional hurdle in applying the resulting knowledge to human-AI collaboration lacks the annotated data~\cite{10.1145/3491102.3502030, DBLP:journals/tvcg/WuGHCRK24}. LLMs present a potential solution by using zero-shot or few-shot learning for rapid prototyping with newly discovered knowledge~\cite{brown2020language, radford2019language}. Our work exemplifies this paradigm by integrating design guidelines into LLMs for AR-specific text simplification.

\subsection{Limitations and Future Work}
Our system uses OpenAI's GPT-3 API, which has a latency of approximately 2 seconds. This latency can result in object positioning inaccuracies if the user moves or turns around. It can also pose challenges in time-sensitive scenarios, such as emergency medical support (E1). Future work could look into device-side post-processing for faster performance. 
In terms of AR devices, we primarily tested the HMD format, thus it is not yet clear if our system will work for other AR devices~\cite{DBLP:conf/uist/QianMLALTH019,DBLP:conf/uist/HartmannY020}.

\section{Conclusion}

In conclusion, this paper presents \systemname, an automated text simplification system tailored for head-mounted AR devices. We first identify the challenges in AR text presentation via a formative study that includes a survey of the literature, an open-ended exploration with seven participants, and interviews with three experts. The findings lead to design guidelines that help form the \systemname~ system. The system leverages OpenAI's GPT-3 models through few-shot learning for automated text simplification. Using chain-of-thought prompting, we present two novel techniques tailored for AR text simplification: a plan-of-technique and error-aware calibration to ensure meaning preservation. We validate our system via a 16-participant empirical study, resulting in significant improvements in users' performance, reduced cognitive load, and better subjective ratings when compared to unmodified text, the state-of-the-art T-5 language model, and other methods. These findings underscore the efficacy of our system in enhancing text readability and mitigating cognitive load during task guidance in AR environments.


\begin{acks}
This work was supported by the DARPA PTG program. Any opinions, findings, and conclusions or recommendations expressed in this material are those of the authors and do not necessarily reflect the views of DARPA. We thank Xiaoan Liu, Jeff Huang, James Tompkin, Leslie Welch, and Ashley Girty for their intellectual contribution and insightful feedback.
\end{acks}

\balance
\bibliographystyle{ACM-Reference-Format}
\bibliography{sample-base}

\appendix

\section{Example Text Simplification Results}
Here, we provide the sample results recorded during our experiments. The Study 1 results are listed in Table~\ref{tab:study1-task1-samples} (Task 1) and Table~\ref{tab:study1-task2-samples} (Task 2).

\begin{table*}[h]
\centering
\begin{tabular}{|r|p{6cm}|p{6cm}|}
\hline
Step & Original  & Simplification    \\
 \hline
1                          & To create a coffee, first please carefully place the pour-over dripper over the coffee mug.                                                                                                                  & { Place dripper (on your left) on coffee mug.}                                            \\ \hline
2                          & Prepare the filter insert by folding the paper filter in half to create a semi-circle, and in half again to create a quarter-circle. Place the paper filter in the dripper and spread open to create a cone. & { Fold paper filter in half, then half again. Put filter in dripper, form cone shape.}    \\ \hline
3                          & Rinse the filter. Pour enough hot water into the filter to wet it. The entire paper filter should be moist. Rinsing the filter will remove any papery residue so your coffee doesn't have a woodsy taste.    & { Wet filter with water to rinse away residue.}                                           \\ \hline
4                          & Lift up the dripper and pour out the water. Then set the dripper with the wet filter back on the coffee mug.                                                                                                 & { Remove dripper, pour out water, and return dripper to coffee mug.}                      \\ \hline
5                          & Get out a digital scale and measure out 3 tablespoons (about 30 g) of coffee beans. Measure out 30 g of whole beans and place them in your grinder.                                                          & { Measure 30g coffee beans on a digital scale (right side), place in grinder (right side).} \\ \hline
6                          & Grind the beans until the coffee grounds are the consistency of coarse sand, about 20 seconds.                                                                                                               & { Grind beans for 20 seconds, until coarse sand consistency.}                             \\ \hline
7                          & Transfer the coffee grounds to the filter cone. Then place the coffee mug with the dripper on a digital scale and set it to zero.                                                                            & { Move grounds to filter cone. Set coffee mug with dripper on scale, zero it.}            \\ \hline
8                          & Slowly pour the water over the grounds in a circular motion. Do not overfill beyond the top of the paper filter. Your scale should read 100 g once you've poured enough water into the dripper.              & { Slowly pour water in circles over grounds, stopping at 100g on scale.}                  \\ \hline
9                          & Let the coffee drain completely into the mug and wait for 30 seconds and you can complete the task;                                                                                  & { Drain coffee into mug and wait for 30 seconds to end.}                                  \\ \hline
\end{tabular}
\caption{Study 1 Task 1 sample result. The original and simplified versions of the text are listed. The sample is collected from P10's experiment session and the simplified results for other participants may vary slightly due to spatial context.}
\label{tab:study1-task1-samples}

\end{table*}

\begin{table*}
\centering
\begin{tabular}{|r|p{6cm}|p{6cm}|}
\hline
Step & Original  & Simplification  \\ 
\hline
1  & Before arranging the meeting room, take a moment to tidy up the desk and move anything that's not necessary to other desks;  & Tidy desk, move the unnecessary items to other desks. \\
 \hline
2  & Once the desk is clear, bring the power strip on the desk and connect the Charger to the power strip so the meeting attendants can use.                                                                           & Put power strip on desk, connect phone charger to it. \\ 
\hline
3 & Connect the camera's charger to the power strip and position the camera at the opposite end of the desk from the TV.    & Connect camera to strip, facing opposite of TV. \\
 \hline
4 & Arrange the chairs in the meeting room. Make sure that there's enough space between each chair - roughly 1.5 feet should suffice. Position one chair on the window side, and place five chairs on the other side. & Arrange chairs on two sides. Leave space of roughly two A4 papers' length apart. Window side: 1 chair. Other side: 5 chairs. \\
 \hline
5  & Next, place cups of water and papers on each chair. Each person should have one cup of water and paper;  & Place water, paper onto desk in front of chairs.\\ 
\hline
6 & Put up the desk nameplates on on each chair. When Alice is on the side of the window, other desk nameplates should be put on the other side. The sequence is Bob, Amy, Andy, Dave and Luis.                       & Place nameplates: Window side: Alice (window); sequence (left to right) on other side: Bob, Amy, Andy, Dave, Luis.\\
 \hline
7 & Since Alice is the VIP in the meeting, place make it clearly by putting the remote controller to Alice’s position.  & Place remote controller at Alice’s position on desk.\\ 
\hline
\end{tabular}
\caption{Study 1 Task 2 sample result. The original and simplified versions of the text are listed. The sample is collected from P10's experiment session and the simplified results for other participants may vary slightly due to spatial context.}
\label{tab:study1-task2-samples}
\end{table*}

\end{document}